\documentclass[useAMS,usenatbib]{mn2e} \input{epsf}

\newif\ifAMStwofonts 
 
\usepackage{longtable}
\usepackage{lscape}
\usepackage[none]{hyphenat} 

\def\lesssim{\mathrel{\hbox{\rlap{\hbox{\lower4pt\hbox{$\sim$}}}\hbox{$<$}}}}
\def\gtrsim{\mathrel{\hbox{\rlap{\hbox{\lower4pt\hbox{$\sim$}}}\hbox{$>$}}}}
\def\msun{${\rm M}_{\odot}$~}
\def\msol{${\rm M}_{\odot}$}

\def\l_lsun{$\log{L/\rm L_{\odot}}$~}
\def\masa_msun{$M/ \rm M_{\odot}$~}

\def\m_mstar{$M/M_{*}$~}

\def\aap{A\&A}

\def\apj{ApJ}
\def\apjl{ApJ}
\def\apjs{ApJS}
\def\mnras{MNRAS}

\def\physrep{Phys. Rep.}

\def\nat{Nature}

\title[Evolution of close binary systems]{The evolution of low mass, 
close binary systems with a neutron star component: a detailed grid}

\author[M.~A.   De   Vito \& O.~G. Benvenuto] {
M.~A.  De Vito$^{1,2}$\thanks{Member of  the Carrera del Investigador
Cient\'{\i}fico,  Consejo Nacional de Investigaciones Cient\'\i ficas y 
T\'ecnicas (CONICET).  Email: adevito@fcaglp.unlp.edu.ar}, 
O.~G. Benvenuto$^{1,2}$\thanks{Member of  the Carrera del Investigador
Cient\'{\i}fico,  Comisi\'on de Investigaciones Cient\'{\i}ficas de la
Provincia de Buenos Aires (CIC).  Email: obenvenuto@fcaglp.unlp.edu.ar}
\\
$^{1}$ Facultad de Ciencias Astron\'omicas y Geof\'{\i}sicas, Universidad
Nacional de La Plata  (UNLP), \\  Paseo del Bosque S/N, B1900FWA, La
Plata, Argentina\\ $^{2}$ Instituto de Astrof\'{\i}sica de La Plata,
IALP, CCT-CONICET-UNLP, Argentina }

\begin{document} 

\date{December 27, 2011}

\pagerange{\pageref{firstpage}--\pageref{lastpage}} \pubyear{2011}

\maketitle \label{firstpage}

\begin{abstract} 

In  close binary  systems  composed of  a  normal, donor  star and  an
accreting  neutron  star,  the  amount  of material  received  by  the
accreting component  is, so  far, a real  intrigue. In  the literature
there are  available models that  link the accretion  disk surrounding
the neutron star with the amount of material it receives, but there is
no model linking  the amount of matter lost by the  donor star to that
falling onto the neutron star.

In  this paper  we explore  the evolutionary  response of  these close
binary systems  when we  vary the amount  of material accreted  by the
neutron star.   We consider a  parameter $\beta$ which  represents the
fraction of material  lost by the normal star that  can be accreted by
the  neutron  star.   $\beta$  is considered  as  constant  throughout
evolution.   We  have  computed  the  evolution of  a  set  of  models
considering initial  donor star  masses $M_{i}$/\msun between  0.5 and
3.50,  initial  orbital periods  $P_{i}$/days  between  0.175 and  12,
initial  masses of  neutron  stars $(M_{\rm  NS})_{i}$/\msun of  0.80,
1.00, 1.20 and  1.40 and several values of  $\beta$.  We assumed solar
abundances.  These  systems evolve to  ultracompact or to  open binary
systems, many of which form  low mass helium white dwarfs.  We present
a grid of calculations and analyze how these results are affected upon
changes in  the value  of $\beta$.  We find a  weak dependence  of the
final donor star  mass with respect to $\beta$. In  most cases this is
also true for the final orbital period. The most sensitive quantity is
the final mass of the accreting neutron star.

As we  do not know the initial  mass and rotation rate  of the neutron
star of  any system, we  find that performing evolutionary  studies is
not helpful for determining $\beta$.

\end{abstract}


\section{Introduction} \label{sec:intro}

It is currently accepted that  close binary systems (CBSs) composed of
a white  dwarf (WD) and a  millisecond pulsar (MSP) are  the result of
the evolution  of a normal, main  sequence donor star  together with a
rotating neutron  star (NS).  These  systems, also, are  considered to
give rise  to the occurrence of  low mass X-ray  binary (LMXB) sources
(see, e.g., Podsiadlowski, Rappaport \& Pfahl 2002).

The  standard model states  that when  a normal  star fills  its Roche
lobe, starts to transfer mass  to its NS companion. The material forms
an  accretion  disk  around  the  compact  star  and  a  part  of  the
transferred mass is deposited onto  the NS surface. The NS rotation is
accelerated  due  to  angular  momentum  deposition  on  its  surface,
becoming a MSP (for a review see Bhattacharya \& van den Heuvel 1991).
In  order to  compute  the evolution  of  CBSs we  have  to make  some
hypotheses on the characteristics  of the mass transfer. Usually, this
problem    has   been    handled   considering    a    two   parameter
description. These  are the  fraction of mass  lost by the  donor star
that  can be accreted  by its  companion ($\beta$)  and the  amount of
specific angular momentum carried out from the system ($\alpha$). Both
quantities  are  assumed  as   constants  during  the  entire  stellar
evolution. The value  of $\beta$ has been usually  set to $\beta= 0.5$
(Podsiadlowski,   Joss  \&   Hsu  1992;   Tauris  \&   Savonije  1999;
Podsiadlowski et al. 2002; Nelson \& Rappaport 2003). In some cases it
has  been set  to $\beta=  1$ ($\beta=  0$) which  represents  a fully
conservative  (non conservative) situation  (Ergma, Sarna  \& Antipova
1998).   Other  values  of  $\beta$  have been  considered  to  fit  a
particular   binary   system   (Benvenuto,   Rohrmann   \&   De   Vito
2006). Meanwhile, it has been usual to set $\alpha= 1$.

The knowledge  of $\beta$ is,  in principle, important for  the binary
evolution models. Its value directly  determines the rate of change of
the NS mass, which affects the  mass ratio, and then the radius of the
Roche lobe  $R_L$ (Eggleton 1983). $\beta$ enters  in the differential
equation that determines the  evolution of the orbital semiaxis which,
in turn, determines the size of the Roche lobe.  Thus, $\beta$ affects
the occurrence of Roche lobe episodes. However, unfortunately, neither
observational evidence  nor theoretical models  allow us to  infer how
much  of  the  matter lost  by  the  donor  star  is accreted  by  the
NS. Besides, there  is an upper limit for the  accretion rate given by
the    Eddington    accretion    rate   $\dot{M}_{Edd}=    2    \times
10^{-8}$~\msol/yr.  To date  $\beta$  has been  considered  as a  free
parameter.   Now, we  may ask  a  question. Can  $\beta$ be  estimated
studying the overall evolution of  these CBSs, as well as its temporal
evolution? To look for the answer is the main aim of this paper.

It  is   know  that  the  appearance  and   variability  of  accreting
millisecond X-ray  pulsars strongly depend on the  accretion rate onto
the NS, $\dot{M}_{\rm NS}$, the 
effective viscosity and diffusivity of
the disk  magnetosphere boundary.  For a  typical NS with  a period of
rotation of  2.5~msec, Romanova et  al.  (2008) present  the following
classification  of  accreting  NSs  as  a  function  of  $\dot{M}_{\rm
  NS}$. At the  {\it boundary layer} regime, if  the accretion rate is
sufficiently large ($\dot{M}_{\rm NS} > 7.3 \times 10^{-8}$~\msol/yr),
the  star's magnetic  field  is completely  buried  (screened) by  the
accreting  matter  that  falls  onto  the star  directly  through  the
boundary layer.  As the accretion rates  decreases, the role played by the
stellar  magnetic field  becomes  more important,  so that it influences
the flow of matter around  the  star.   When  the  mass  transfer  rate  is
sufficiently low ($1.3 \times 10^{-11}$~\msol/yr $< \dot{M}_{\rm NS} <
1.4 \times 10^{-9}$~\msol/yr)  the magnetosphere radius becomes larger
than the  corotation radius, and  the star enters the  {\it propeller}
regime. In  the strong propeller regime, disk  matter acquires angular
momentum from the  rotating magnetosphere fast enough that  most of it
is  ejected by  a conical  outflow. At  the same  time,  a significant
amount  of angular  momentum and  energy flow  along the  open stellar
field lines, giving axially symmetric jets.  Finally, for even smaller
accretion rates  ($\dot{M}_{\rm NS} <  1.3 \times 10^{-11}$~\msol/yr),
accretion onto  the NS surface is  suppressed, and the  star becomes a
pulsar.  This is  the {\it  pulsar}  regime.  Evidently,  there is  an
important relation between the magnetic  field intensity of the NS and
its accretion rate.   The above given values for these regimes
should increase for a stronger  magnetic field.  Thus, to find $\beta$
we would need to compute the NS magnetic field evolution.

In  the standard  model of  accreting NSs  (or black  holes,  BH), the
system NS(BH)-accretion  disk is considered (Shakura  \& Sunyaev 1973;
White, Stella \&  Parmar 1988; Mitsuda et al.   1989; Church, Inogamov
\& Baluci\'nska-Church et al.   2002; Kulkarni \& Romanova 2009).  The
structure and  radiation of stationary disks around  NSs is determined
by several  parameters: the  mass of the  NS, the accretion  rate, the
level of turbulence and/or small scale magnetic fields, etc. If matter
flows through the  inner boundary at a rate  substantially higher than
$\dot{M}_{Edd}$,  the gas  should flow  away perpendicularly  from the
inner region of  the disk driven by radiation  pressure.  Many authors
have developed models of two  and three components in order to account
for  the  observed emission  spectra  in  these NS(BH)-accretion  disk
systems.  However, these models do  not link the amount of matter lost
by the donor star that is accreted by the NS(BH).

Takahashi \& Makishima (2006) show that the energy spectra of 18~LMXBs
is successfully accounted for by  a model consisting of a canonical NS
($M_{\rm     NS}=$~1.40~\msol)     with     $\dot{M}_{\rm    NS}     <
\dot{M}_{Edd}$.  They consider  a combination  of two  optically thick
components, one  due to the accretion  disk and the  other radiated by
the NS surface.  As the accretion rate increases,  the disk luminosity
increases  but the  emission from  the  NS surface  saturates or  even
decreases.   When $\dot{M}_{\rm NS}  \gtrsim \dot{M}_{Edd}$,  the LMXB
spectrum  consists of  three optically  thick components;  the softest
from  a  retreated disk,  the  hardest from  the  NS  surface, and  an
intermediate component  presumably due to  the outflows caused  by the
increased radiation pressure.

Again, we could  establish a link between the type  of model that fits
the energy  spectrum of these  objects and $\dot{M}_{\rm  NS}$.  Then,
according to  the best fit to the  energy spectrum of the  NS we could
model the  accretion onto the  NS and consequently model  $\beta$.  In
any case,  the value of $\beta$  found in this way  corresponds to the
very short timescale of observations, while in evolutionary studies we
need the value of $\beta$ averaged on far longer time periods.

Evidently, computing $\beta$ from first principles is a very difficult
task.  It may  be considered that a way to find  $\beta$ is to compute
the effects  on the evolution of  CBSs induced by  changes in $\beta$.
In order to  explore the viability of such strategy,  in this paper we
compute  a  grid  of   evolutionary  models.   We  consider  the
evolution of solar composition donor  stars members of CBSs for a wide
range of  initial parameters  (masses for the  donor star  $M_{i}$ and
accreting NS $(M_{\rm NS})_{i}$,  and orbital periods $P_{i}$).  Also,
we  consider  different values  for  $\beta$~(between  0  and 1,  with
$\Delta    \beta   =    $~0.25)    for   the    cases   of    $(M_{\rm
  NS})_{i}$/\msol=~0.80,  1.00  and 1.20,  and  with  $\Delta \beta  =
$~0.125 for  the case of $(M_{\rm  NS})_{i}$/\msol=~1.40 extending our
previous  calculations (with  $\beta$= 0.5)  presented in  De  Vito \&
Benvenuto  (2010).  For  simplicity,  we shall  consider that  $\beta$
remains constant  along each calculation.  Then, we shall  analyze the
sensitivity  of the  evolutionary tracks due to changes  in  $\beta$.  We
shall be particularly  interested in helium WDs, that  are expected to
be  the  type  of  objects  found  in some  CBSs  with  accurate  mass
determinations (see below, \S~\ref{sec:discu}).

The   reminder   of   the   paper   is  organized   as   follows:   in
Section~\ref{sec:thecode} we  present the main  characteristics of our
evolutionary code. In Section~\ref{sec:results} we present and analyze
the results obtained from our calculations. The main part of the paper
ends in  Section~\ref{sec:discu} where we  discuss of our  results and
make some  concluding remarks.  In  Appendix~\ref{app_grid} we present
tables of our main numerical  results and the relation between WD mass
and the final orbital period is given in Appendix~\ref{app_mwdp}.

\section{The computer code} \label{sec:thecode} 

The  code employed here  has been  presented in  Benvenuto \&  De Vito
(2003) where we described a  generalized algorithm based on the Henyey
technique that  allows for the  simultaneous computation of  the donor
stellar  structure and  the mass  transfer  rate in  a fully  implicit
way. The code has updated physical ingredients. For temperatures ${\rm
  T} >  6 \times  10^3$~K we considered  radiative opacities  given by
Iglesias  \& Rogers  (1996) while  at lower  temperatures  we employed
molecular  opacities given  by  Ferguson et  al.  (2005).   Conductive
opacities have  been taken  from Itoh et  al. (1983). Our  equation of
state has been  that of Magni \& Mazzitelli  (1979).  Nuclear reaction
rates  have  been taken  from  Caughlan  \&  Fowler (1988).   Neutrino
emission has  been described  following the works  by Itoh  \& Kohyama
(1983); Munakata, Kohyama \& Itoh  (1987); Itoh et al. (1989) and Itoh
et al. (1992).   Diffusion processes (gravitational settling, chemical
and thermal  diffusion) have been  accounted for following  Althaus \&
Benvenuto (2000). We consider the Mixing Length Theory as described in
Kippenhahn, Weigert  \& Hofmeister  (1967), setting the  Mixing Length
parameter $l$  to $l/H_p=  1.7432$ (here $H_p$  is the  pressure scale
height defined by $H_p \equiv  dr/d\ln{P}$ where $P$ denotes the total
pressure and  $r$ is the  distance measured from the  stellar centre).
Convective core overshoot  is included as in Demarque  et al.  (2004).
This important physical phenomenon consist in the presence of material
motions  and  mixing  beyond  the canonical  boundary  for  convection
defined by  the clasic Schwarzschild criterium. A  proper treatment of
convective  core  overshoot  would  require a  radiative  hydrodynamic
treatment near the convective edge.  The overshoot lenght is evaluated
in terms of the local  pressure scale height, multiplied by a constant
parameter less  than unity ($\Lambda_{OS}$). In  their paper, Demarque
et al. (2004) use values of  $\Lambda_{OS}$ from 0 to 0.2 depending on
the  value of  the  stellar mass  compared  to $M^{conv}_{crit}$  (the
critical  mass above which  stars have  a substantial  convective core
after pre  main sequence  phase). This value  depends on  the chemical
composition.  For   futher  details  see  Demarque   et  al.   (2004).
Furthermore,  we considered  grey atmospheres  and  neglected external
irradiation due to the companion.

Let  us now  quote the  physical  ingredients we  considered that  are
specifically related to binary evolution.  To compute the radius $R_L$
of a sphere with a volume equal to that of the Roche Lobe, we employed
the standard expression given by  Eggleton (1983). We adopted the mass
transfer rate expression given by Ritter (1988). The orbital evolution
has  been computed  following Rappaport,  Joss \&  Webbink  (1982) and
Rappaport, Verbunt  \& Joss (1983).  Mass and angular  momentum losses
have  been  described by  two  free  parameters  $\alpha$ and  $\beta$
(defined  above). Gravitational  radiation and  magnetic  braking were
described as in Landau \&  Lifshitz (1975) and Verbunt \& Zwaan (1981)
respectively.

In  our  treatment of  the  orbital  evolution,  as stated  above,  we
consider  that the  NS is  able to  retain a  $\beta$ fraction  of the
material  coming   from  the  donor  star   $\dot{M}_{\rm  NS}=  \beta
|\dot{M}|$ (where $\dot{M}$  is the mass transfer rate  from the donor
star\footnote{We  use   absolute  value  because,   according  to  our
  definition $\dot{M}$ is a negative quantity.}), as done in Benvenuto
\&  De  Vito (2005).   We  considered  that  $\beta$ remains  constant
throughout all Roche lobe  overflow (RLOF) episodes.  Also, we assumed
that material lost  from the binary systems carries  away the specific
angular momentum of the compact object ($\alpha= 1$).

\section{Numerical results} \label{sec:results}

We  have constructed  a  grid  of evolutionary  models  for the  donor
component  of CBSs.   We considered  a  wide range  of initial  masses
$M_{i}$ for  the normal, solar metallicity  star ($M_{i}$/\msun= 0.50,
0.65,  0.80, 1.00,  1.25, 1.50,  1.75, 2.00,  2.25, 2.50,  2.75, 3.00,
3.25, 3.50).  For the mass  $(M_{\rm NS})_{i}$ of the accreting NS, we
have    selected    four    different   initial    values    ($(M_{\rm
  NS})_{i}$/\msun=~0.80, 1.00, 1.20 and 1.40).

While most of the known NS  masses are around 1.4~\msol, some NSs have
masses  clearly below that  value. Good  examples are  the NSs  in the
X-Ray  binaries  SMC  X-1,  Cen  X-3  and  4U1538-52  with  masses  of
$1.17^{+0.16}_{-0.16}$,           $1.09^{+0.20}_{-0.36}$,          and
$0.96^{+0.19}_{-0.16}$~\msun  respectively (Lattimer \&  Prakash 2004;
Lattimer \& Prakash  2007). This justifies our choice  of 0.8~\msun as
the minimum  value for $(M_{\rm NS})_{i}$. Very  recently, Demorest et
al. (2010)  have detected  a NS  with a mass  of $M_{\rm  NS}=1.97 \pm
0.04$\msun in the PSR J1614-2230 binary system (with an orbital period
of  8.6866194196~days) orbiting  together with  a WD  of  $M=0.500 \pm
0.006$\msol. While this detection  indicates that considering NSs with
initial masses larger than 1.4~\msun should also be meaningful, notice
that the WD is too massive  to have a helium rich interior (see, e.g.,
Iben \& Tutukov 1985). Thus,  this binary should not correspond to the
class of systems we are interested in here.

We choose the  initial orbital period of the  systems $P_{i}$ in order
to obtain helium WDs or  members of ultracompact binary systems (those
in which the orbital period is  less than 1~h; see, e. g., Fedorova \&
Ergma 1989; van der Sluys, Verbunt \& Pols 2005) as the final state of
the  donor  stars.  Besides,  for  each  group  of initial  parameters
(masses of the components and orbital periods) we have considered five
values of $\beta$  (0.00, 0.25, 0.50, 0.75 and 1.00)  for the cases of
$(M_{\rm  NS})_{i}$/\msun=~0.80,  1.00,  1.20  and  refined  our  grid
considering a step of $\Delta \beta  = 0.125$ for the case of $(M_{\rm
  NS})_{i}$=~1.40~\msol.  We  have  performed  more  than  a  thousand
evolutionary sequences in which we  have followed the evolution of the
donor star from the ZAMS on.  In order to end our calculations we have
considered several situations. As we  are interested in helium WDs, we
only  consider objects with  a central  temperature $\log_{10}{(T_{\rm
    c}/K)} < 8$, below the threshold for helium burning. Also, we stop
if the mass transfer rate exceeds a value of $10^{-5}$~\msol/yr, or if
the mass of the accreting NS  is greater than 2.5~\msol. This value is
larger  than the  maximum mass  of  NS corresponding  to many  nuclear
matter equations of state (Lattimer  \& Prakash 2004).  In the case of
systems that evolve to  an open configuration, we stopped computations
if the WD luminosity is $\log_{10}{(L/L_{\odot})} \leq -5$ or if it is
much older  (20~Gyr) than the  Universe.  In the case  of ultracompact
systems, we ended the calculations when $M\leq 0.050$~\msun or $P \leq
0.05$~days.

As we  varied the  parameters defining  the CBS over  a wide  range of
values, it  is not surprising  that we have  found a large  variety of
evolutionary paths. In some cases  the mass transfer episode is stable
and  the rate  of mass  exchange  is self-regulated,  while in  others
$|\dot{M}|$  increases  to extreme  values  leading  to common  envelope
evolution.

It is known that a dynamical mass transfer instability occurs when the
radius of the Roche lobe shrinks more rapidly (or expands less slowly)
than the donor star. The adiabatic response of a star to mass loss has
long  been understood (see,  e.g., Hjellming  \& Webbink  1987). Stars
with radiative envelopes (e.g., upper main sequence stars) contract as
response  to mass  loss while  stars with  convective  envelope (e.g.,
lower  main  sequence or  red  giant  stars)  expand (for  a  detailed
explanation  see, e.g.,  Soberman,  Phinney \&  van  den Heuvel  1997;
Podsiadlowski  et al.  2002).  When  the donor  star  is perturbed  by
removal  of  some  mass,  it  falls out  of  hydrostatic  and  thermal
equilibrium.  In  the process of reestablishing  equilibrium, the star
will either grow  or shrink.  Also the Roche  lobe changes in response
to the  mass transfer/loss.   As long as  the donor star's  Roche lobe
continues to enclose  the star, mass transfer is  stable. Otherwise it
is unstable and  proceeds on a dynamical timescale.  We define for the
donor  star and  its Roche  lobe  $\zeta_{donor} =  \partial \ln{R}  /
\partial  \ln{M}$  and $\zeta_{L}  =  \partial  \ln{R_{L}} /  \partial
\ln{M}$ respectively.  The stability of mass transfer is determined by
a comparison  of $\zeta_{donor}$  and $\zeta_{L}$. Given  $R\cong R_L$
(the condition for the onset  of RLOF) the initial stability criterion
is  $\zeta_{L} \leq  \zeta_{donor}$.  Tauris \&  Savonije (1999)  have
studied the behavior of $\zeta_{L}(q,\beta)$ for LMXBs, where $q = M /
M_{\rm NS}$.  They found that  $\zeta_{L}$ does not depend strongly on
$\beta$,   which  is   in   agreement  with   our  calculations   (see
Tables~\ref{table:mns0.80}~-~\ref{table:mns1.40}). These authors found
that,  in general,  the Roche  lobe increases  ($\zeta_{L} <  0$) when
material is transferred  from a light donor to a heavier  NS ($q < 1$)
and correspondingly  $R_L$ decreases ($ \zeta_{L} >  0$) when material
is transferred from  a heavier donor to a lighter NS  ($q > 1$). These
are  the cases  (a large  value of  $q$ and/or  stars  with convective
envelope) where we find unstable mass transfer situations.

On the  contrary, for the  case of the  most massive initial  NSs, our
grid extends to  initial masses of the donor star  up to 3.5~\msol. In
this case the calculations are  stopped because of the onset of helium
burning at  the stellar core  ($\log_{10}{(T_{\rm c}/K)} \geq  8$), or
because the NS mass exceeds the upper limit we have chosen (especially
in the case  of $\beta=$~1). If we consider  higher initial values for
the   mass  of   the  NS,   this   situation  would   be  found   more
frequently.  Presumably,  these  CBSs  should lead  to  BH  formation.
Another interesting  result is that  the range of initial  periods for
which  CBSs lead  to the  formation  of converging  systems extend  to
higher  initial  orbital periods  ($P_i$)  with  decreasing values  of
$(M_{\rm NS})_i$.

\begin{figure*}  
\epsfysize=200pt 
\epsfbox{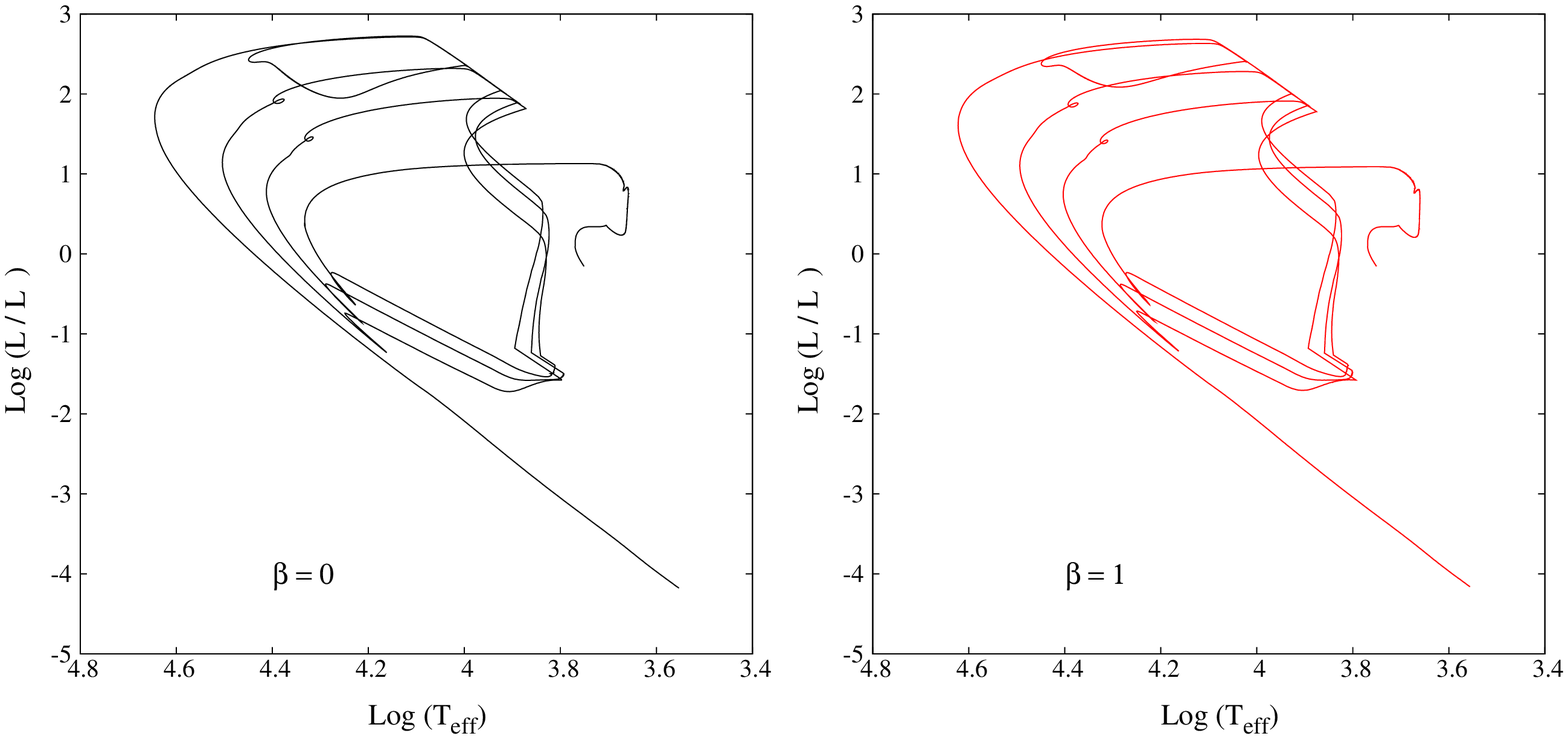}  
\epsfysize=200pt 
\epsfbox{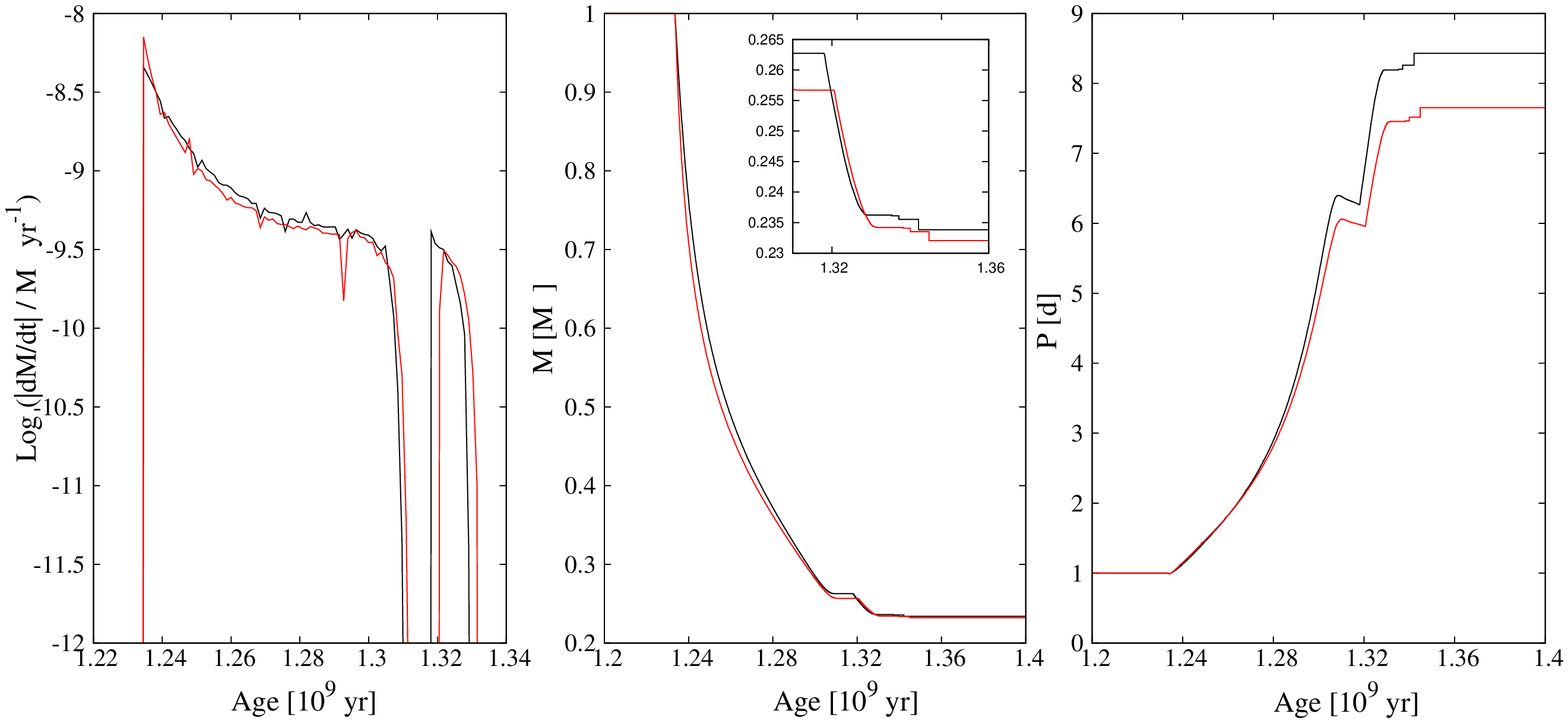} 
\caption{\label{fig:hr_p_mdot_m}  Upper panels  show  the evolutionary
  tracks  of  the  donor  component   of  a  CBSs  with  initial  mass
  $M_i=$~1~\msun for the  donor star, $(M_{\rm NS})_i=$~1.40~\msun for
  the   neutron    star,   and   an   initial    orbital   period   of
  $P_{i}=$~1.5~days. The three loops in the H-R diagrams are due 
  to hydrogen shell flashes and very little mass transfer is associated 
  with any beyond the first mass transfer episode (for details, see main
  text). Left  (right) panel  corresponds to the  case of
  $\beta=$ 0.0 (1.0).  Lower  panels show the results corresponding to
  the same  evolutionary calculations related to the  evolution of the
  mass transfer  rate during the  first RLOF (left panel),  donor mass
  (middle panel)  and orbital period  (right panel).  Notice  that the
  final period is  slightly dependent on $\beta$ while  the others are
  almost unaffected.} 
\end{figure*}

In  order to  analyze the  changes in  the evolution  of open  CBSs by
varying  $\beta$, we  may choose  the set  of models  corresponding to
donor  stars with initial  mass $M_i=$~1.00~\msol,  a NS  with initial
mass   $(M_{\rm  NS})_i=$~1.40~\msol,   initial   orbital  period   of
$P_{i}=$~1.5~days  and  extreme values  of  $\beta$  (0  and 1)  as  a
representative     case.       On     the     upper      panels     of
Fig.~\ref{fig:hr_p_mdot_m} we  present the evolutionary  tracks of the
donor star for these systems.

After core hydrogen exhaustion, the donor star evolves towards the red
giant region  of the HR  diagram, overflowing its  corresponding Roche
lobe.  Since  then, the  star undergoes the  first RLOF  mass transfer
episode. After  losing approximately~$70\%$  of its initial  mass, the
outer hydrogen envelope embraces a  so little mass fraction that it is
no longer  able to stand as a  giant and starts a  fast contraction to
become a  pre~WD star.   This contraction heats  up the bottom  of the
hydrogen  envelope   that  now  is   partially  degenerate,  meanwhile
diffusion  has leaded  some  hydrogen inwards.   Then a  thermonuclear
hydrogen flash starts,  leading to a sudden swell  of the outer layers
that  overflow the  Roche lobe  again. Now  the amount  of transferred
matter is  far lower (approximately~$10^{-3}$\msol) than  that lost by
the donor star during the first RLOF.  This transferred mass, together
with the nuclear  burning, still active at the  bottom of the hydrogen
envelope,  contribute  to lower  the  total  hydrogen  content of  the
star. This  forces the star to  undergo a new contraction  to become a
pre~WD star  again. In  this set of  models, the donor  star undergoes
three flashes before  evolving to the final WD  cooling track.  A more
detailed discussion of the evolution  of this kind of systems has been
presented in Benvenuto \& De Vito (2004).

Notice     that      the     evolutionary     tracks      shown     in
Fig.~\ref{fig:hr_p_mdot_m} are  barely dependent on  $\beta$. The same
is found when  we analyze the evolution of the  mass transfer rate and
the mass  of the donor  star, as shown  in the bottom left  and middle
panels  of Fig.~\ref{fig:hr_p_mdot_m}.  More  significant changes  are
found for the evolution of the  orbital period as shown in the bottom,
right panel of Fig.~\ref{fig:hr_p_mdot_m}.

\begin{figure*} 
\epsfysize=200pt \epsfbox{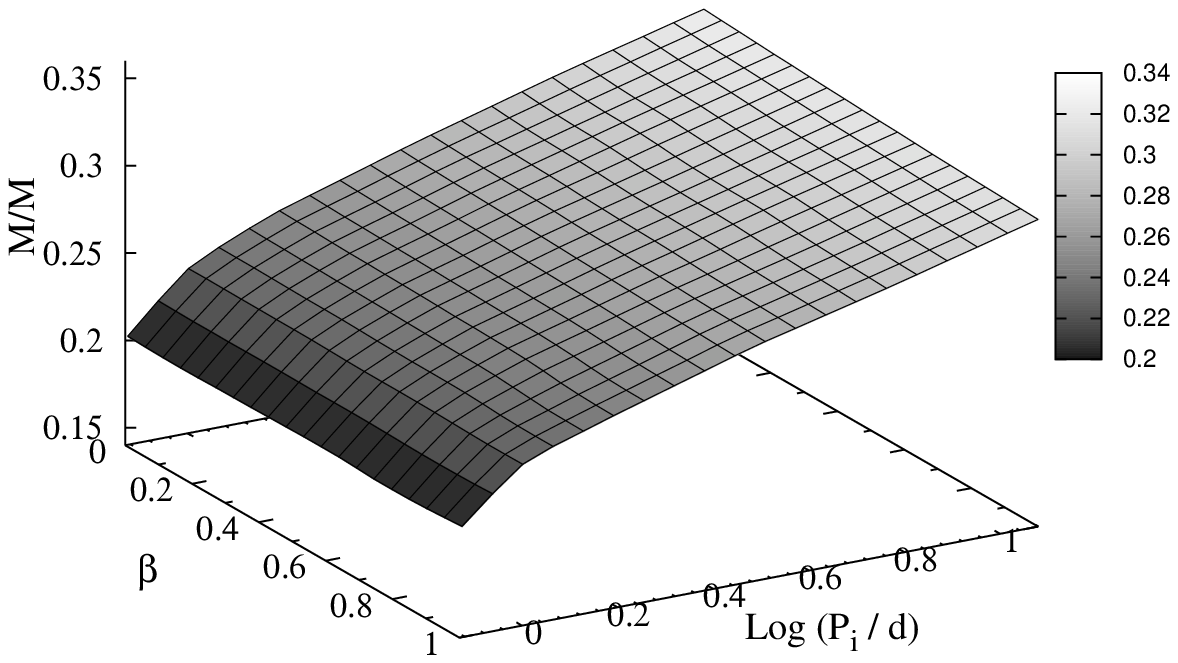}
\epsfysize=200pt \epsfbox{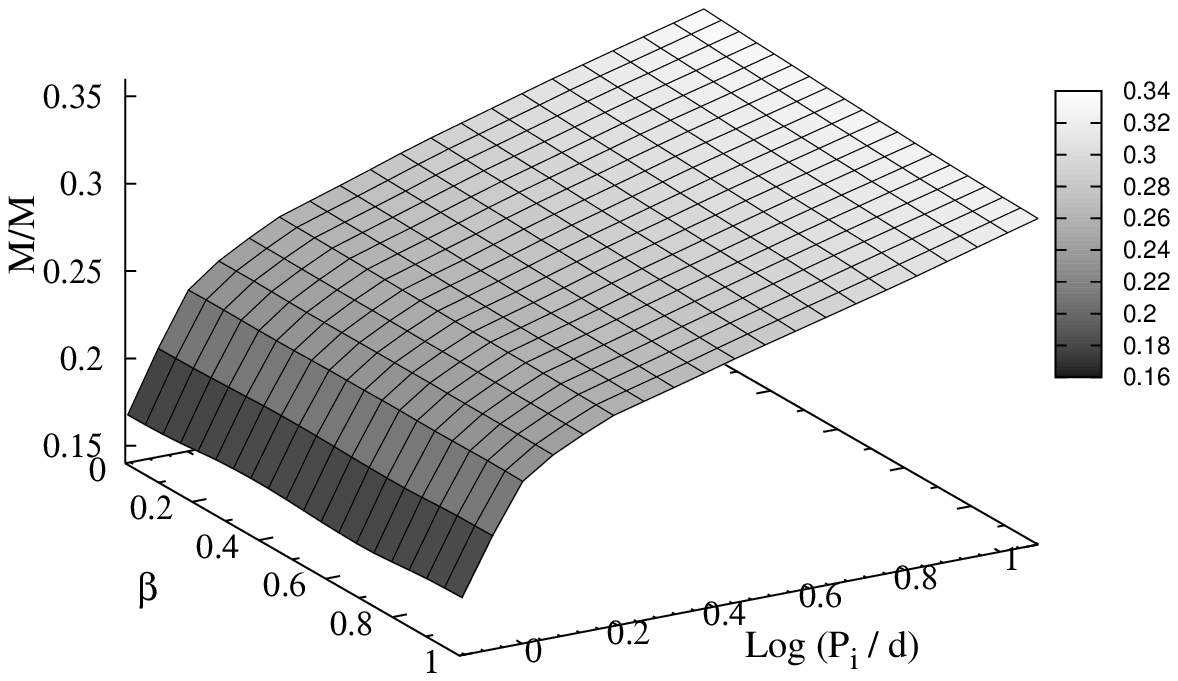}
\epsfysize=200pt \epsfbox{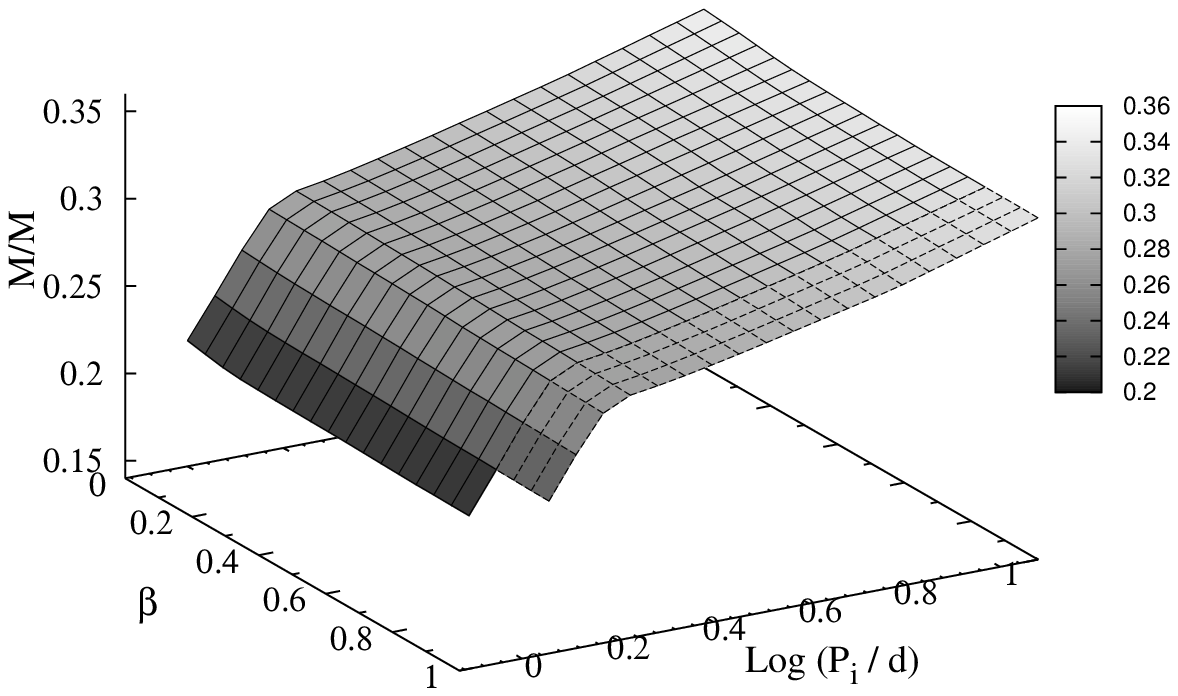}
\caption{The final mass of the donor star for the case of systems with
  $(M_{\rm   NS})_{i}=$~1.4~$M_{\odot}$    and   normal   stars   with
  $M_{i}=$~1.00~$M_{\odot}$  (upper  panel), $M_{i}=$~1.25~$M_{\odot}$
  (middle  panel), and $M_{i}=$~1.50~$M_{\odot}$  (bottom panel)  as a
  function of the logarithm of  the initial orbital period $P_{i}$ (in
  days) and the  fraction $\beta$ of the mass that  can be accreted by
  the NS.  The  surface corresponding to the case  of an initial donor
  mass of  $1.50$~$M_{\odot}$ does not extend on  a rectangular region
  because in the region not shown, the mass of the NS gets larger than
  $2.50$~$M_{\odot}$.   Notice  that the  grey  scale  on the  surface
  indicates the mass values.
  \label{fig:superf_mdonor}}
\end{figure*} 

\begin{figure*} 
\epsfysize=200pt \epsfbox{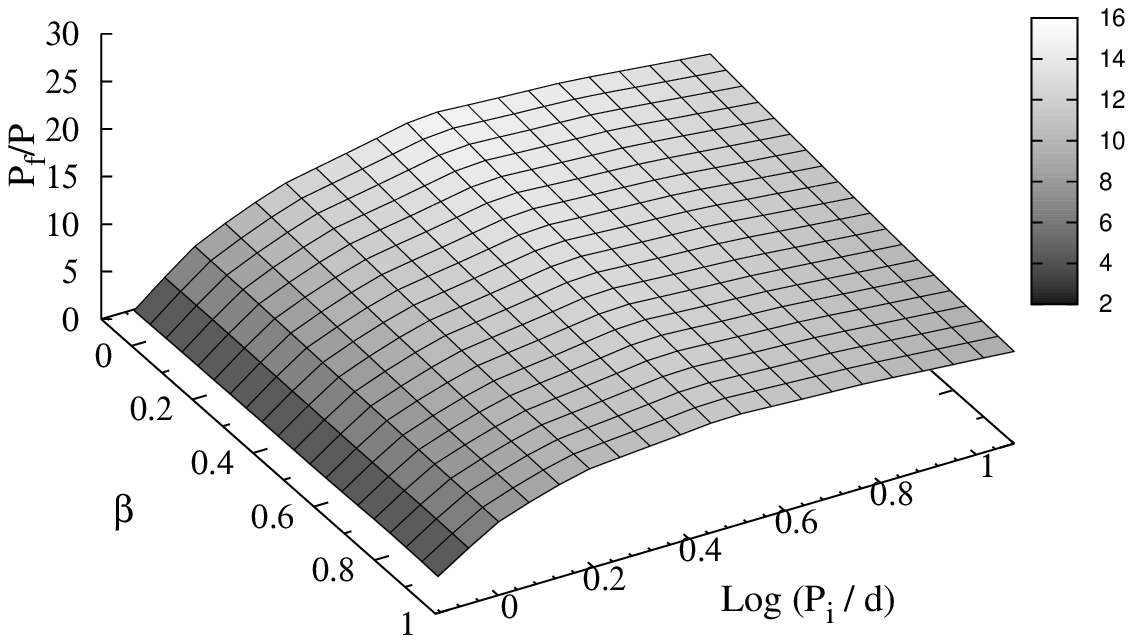}       
\epsfysize=200pt \epsfbox{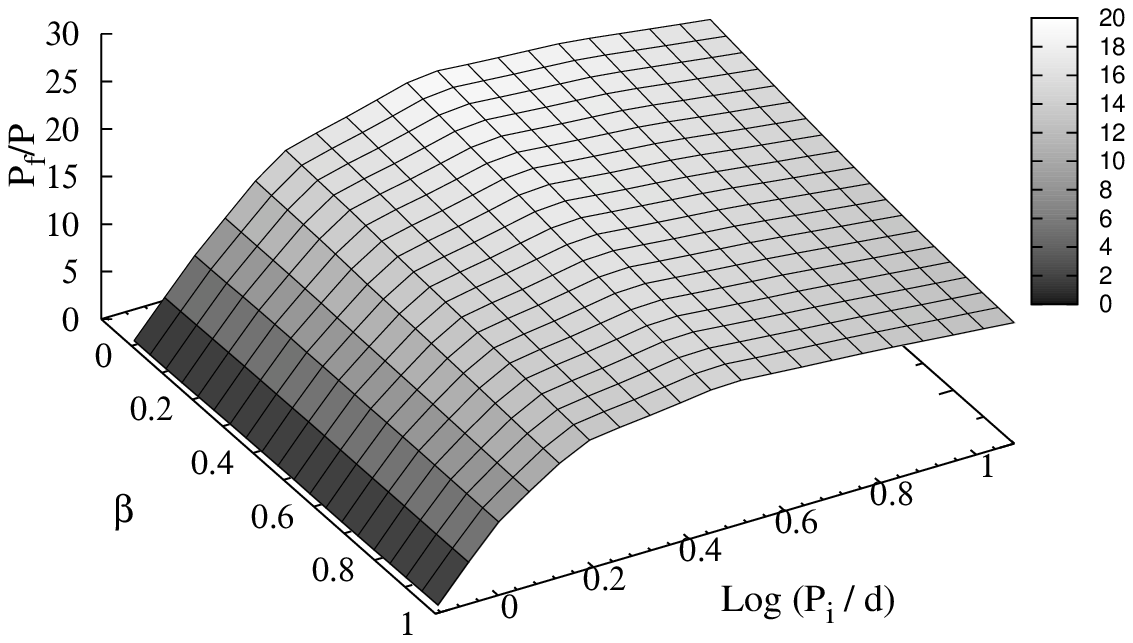} 
\epsfysize=200pt \epsfbox{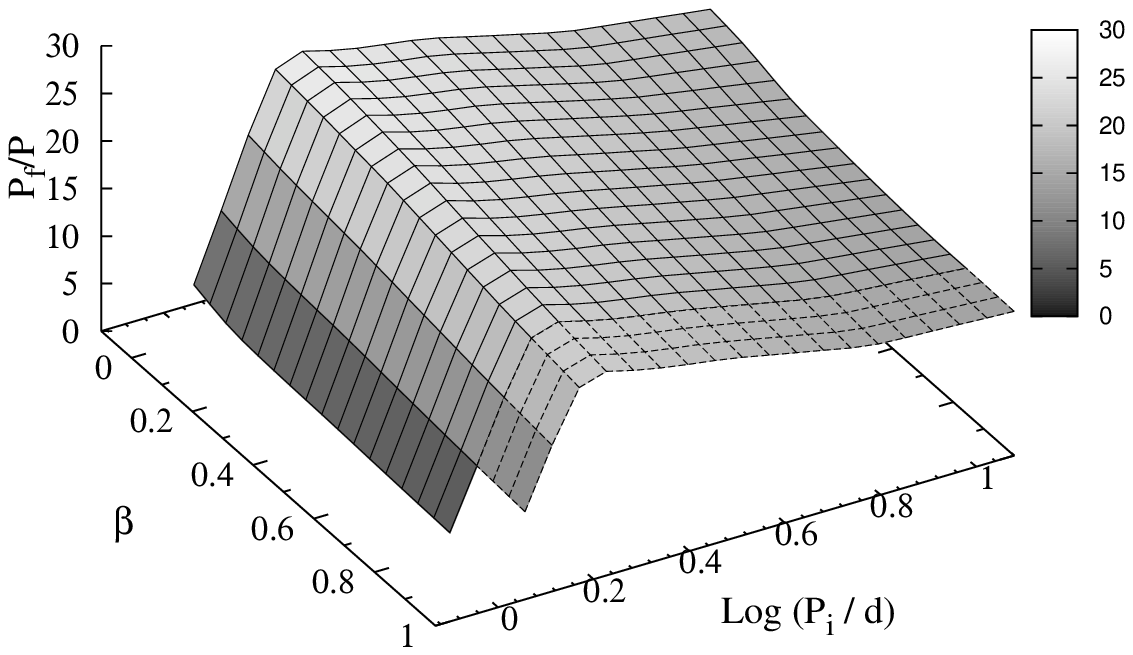}
\caption{The ratio of the final  to the initial orbital period for the
  case  of  systems  with $(M_{\rm NS})_{i}=$~1.4~$M_{\odot}$  and  normal
  stars     with      $M_{i}=$~1.00~$M_{\odot}$     (upper     panel),
  $M_{i}=$~1.25~$M_{\odot}$         (middle         panel),        and
  $M_{i}=$~1.50~$M_{\odot}$  (bottom  panel)  as  a  function  of  the
  logarithm of the initial orbital period $P_{i}$ (in days) and the 
  fraction $\beta$ of the mass that can
  be accreted by the NS.  The  surface corresponding to the case of an
  initial  donor  mass of  $1.50$~$M_{\odot}$  does  not  extend on  a
  rectangular region because in the  region not shown, the mass of the
  NS gets larger than $2.50$~$M_{\odot}$. As in
  Fig.~\ref{fig:superf_mdonor}, the grey scale on the surface corresponds
  to the value of the function on the vertical axis.
  \label{fig:superf_porbital}}
\end{figure*} 

\begin{figure*} \epsfysize=200pt
\epsfysize=200pt \epsfbox{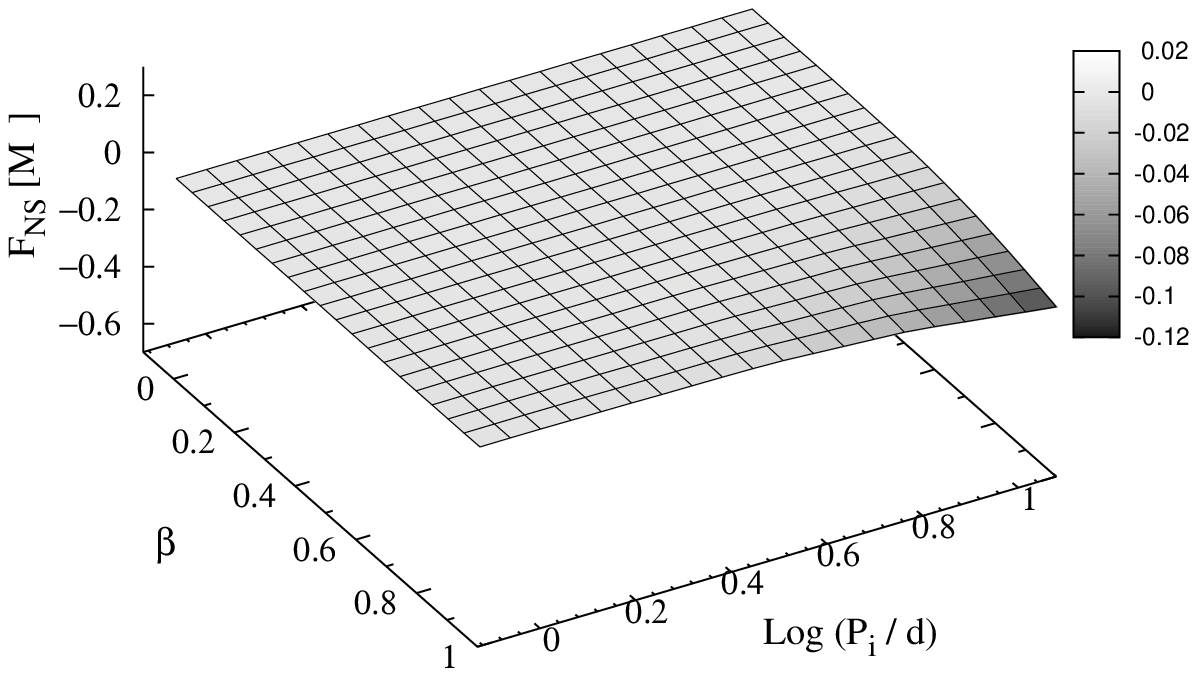} 
\epsfysize=200pt \epsfbox{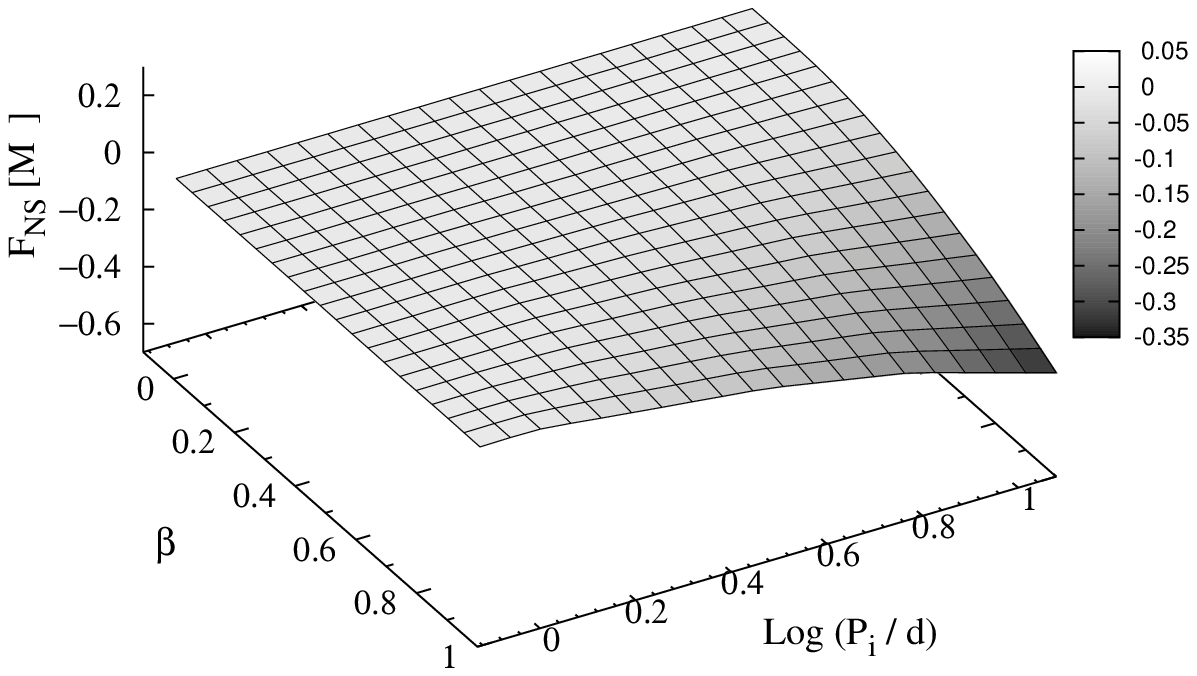} 
\epsfysize=200pt \epsfbox{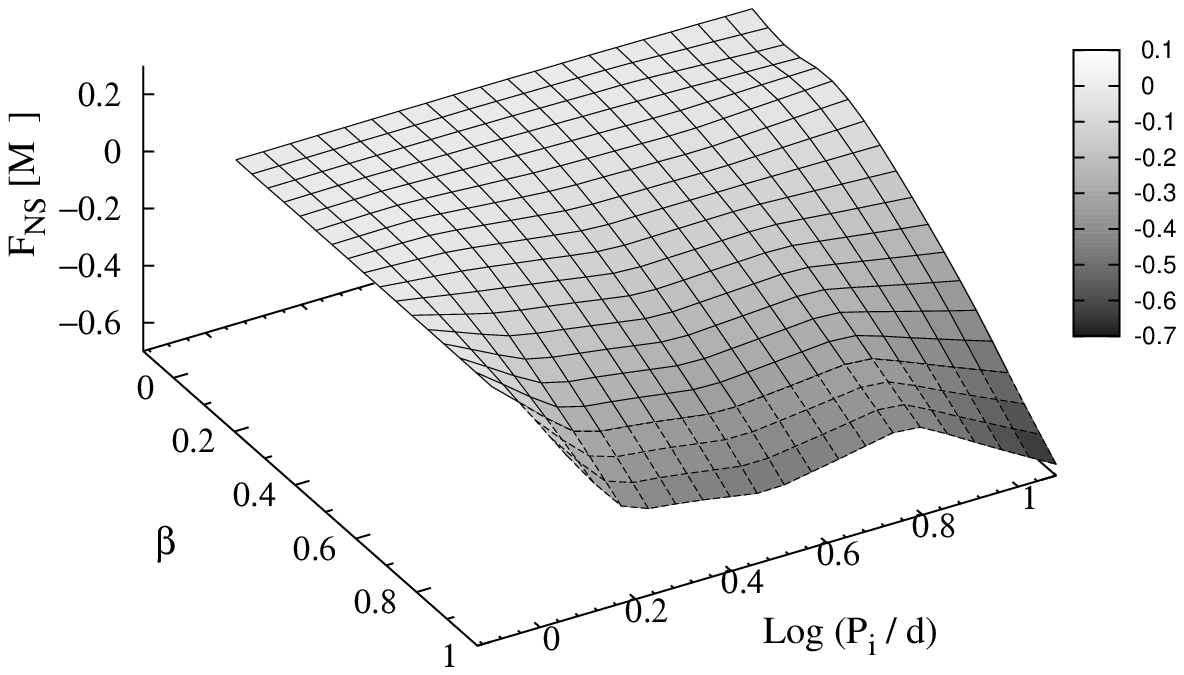}

\caption{The function $F_{\rm NS}= M_{\rm NS}-(M_{\rm NS})_{i} + \beta
  \big(M -  M_{i} \big)$ (defined in  Eq.~\ref{eq:transf_mns}) for the
  case of systems  with $(M_{\rm NS})_{i}=$~1.4~$M_{\odot}$ and normal
  stars     with      $M_{i}=$~1.00~$M_{\odot}$     (upper     panel),
  $M_{i}=$~1.25~$M_{\odot}$         (middle         panel),        and
  $M_{i}=$~1.50~$M_{\odot}$  (bottom  panel)  as  a  function  of  the
  logarithm of  the initial orbital  period $P_{i}$ (in days)  and the
  fraction $\beta$  of the mass  that can be  accreted by the  NS. The
  surface  corresponding to  the  case  of an  initial  donor mass  of
  $1.50$~$M_{\odot}$ does  not extend on a  rectangular region because
  in  the region  not  shown, the  mass  of the  NS  gets larger  than
  $2.50$~$M_{\odot}$.  $F_{\rm NS}$ gives  the amount of material lost
  from  the binary  system because  of super-Eddington  acretion rates
  onto the NS.  If $\dot M_{\rm NS} \le  \dot{M}_{Edd}$ were fulfilled
  during all  RLOFs, then $F_{\rm  NS}= 0$.  The departure  of $F_{\rm
    NS}$  from   zero  is  barely   noticeable  for  the  case   of  a
  $M_{i}=$~1.00~$M_{\odot}$  donor  star.  However,  for the  case  of
  $M_{i}/M_{\odot}=$~1.25  and 1.50 the  surfaces get  larger negative
  values the larger $\beta$ and $P_{i}$.  These conditions corresponds
  to  short RLOF episodes  when the  donor is  a red  giant undergoing
  super-Eddington transfer  rates. As in Fig.~\ref{fig:superf_mdonor},
  the  grey scale  on  the surface  corresponds  to the  value of  the
  function on the vertical axis.
\label{fig:superf_mns} }
\end{figure*} 

\begin{centering}
\begin{table*}
\caption{Values of  the coefficients corresponding  to the fit  of the
  function  defined in  Eq.~(\ref{eq:fit})  to the  surfaces shown  in
  Figs.~\ref{fig:superf_mdonor}-\ref{fig:superf_mns}.   In   the  last
  column we give the maximum relative error of the fit with respect to
  the numerical results.  In most regions of these  surfaces the error
  is much smaller.
\label{table:fits} }
\begin{tabular}{cccccccccc}
\hline
\hline
        &    \multicolumn{8}{c|}{Fit to $M_f$ $[M_{\odot}]$ } \\
$M/M_{\odot}$ & $C_{1}$  & $C_{2}$  & $C_{3}$  & $1000 \times C_{4}$ & $C_{5}$  & $C_{6}$  & $C_{7}$ & $C_{8}$ & Error \\
\hline
1.00 &  0.23300 & -0.034933 & 1.02946 & -0.67879  & 0.240747 & -0.057173 & -0.143095 & 3.70149 & 5\% \\
1.25 &  0.22853 & -0.027247 & 1.20325 & -0.51458  & 0.244226 & -0.064110 & -0.125305 & 4.08725 & 1\% \\ 
1.50 &  0.20805 & -0.031279 & 2.91613 &  7.29667  & 0.463526 & -0.109318 & -0.090219 & 9.69335 & 1\% \\
\hline
\hline
        &    \multicolumn{8}{c|}{Fit to $P_f/P_i$ } \\
$M/M_{\odot}$ & $C_{1}$  & $C_{2}$  & $C_{3}$  & $C_{4}$  & $C_{5}$  & $C_{6}$  & $C_{7}$ & $C_{8}$ & Error \\
\hline
1.00 &  8.48258 &  0.115165 & 41.7855 & 0.502319 & -18.7955 & -6.80065 & 0.212949 & 1.36773 & 5\% \\
1.25 &  8.42498 & -0.317404 & 65.9782 & 0.129753 & -30.7561 & -10.5503 & 0.076967 & 1.5722  & 1\% \\
1.50 &  4.21298 & -2.18037  & 387.409 & 1.42005  & -180.716 & -50.0681 & 0.299583 & 10.352  & 2\% \\
\hline
\hline
        &    \multicolumn{8}{c|}{Fit to $F_{\rm NS}$ $[M_{\odot}]$ } \\
$M/M_{\odot}$ & $100 \times C_{1}$  & $100 \times C_{2}$  & $100 \times C_{3}$  & $100 \times C_{4}$  & $100 \times C_{5}$  & $C_{6}$  & $C_{7}$ & $C_{8}$ & Error \\
\hline
1.00 & -0.45035 & 1.63095 & 1.21402 & -1.35300 & -0.64233 & -0.017375 & -0.454202 & -0.390476 & 10\% \\ 
1.25 & -1.16727 & 6.38160 & 3.61386 & -6.21125 & -1.62558 & -0.105957 & -0.351645 & -0.332994 & 3\% \\  
1.50 & -1.19452 & 3.22075 & 14.5978 & -2.16577 & -4.70089 & -0.730932 & -0.901429 &  1.134020 & 5\% \\ 
\hline \hline
\end{tabular} \end{table*} \end{centering}

\begin{figure*}  \epsfysize=300pt  \epsfbox{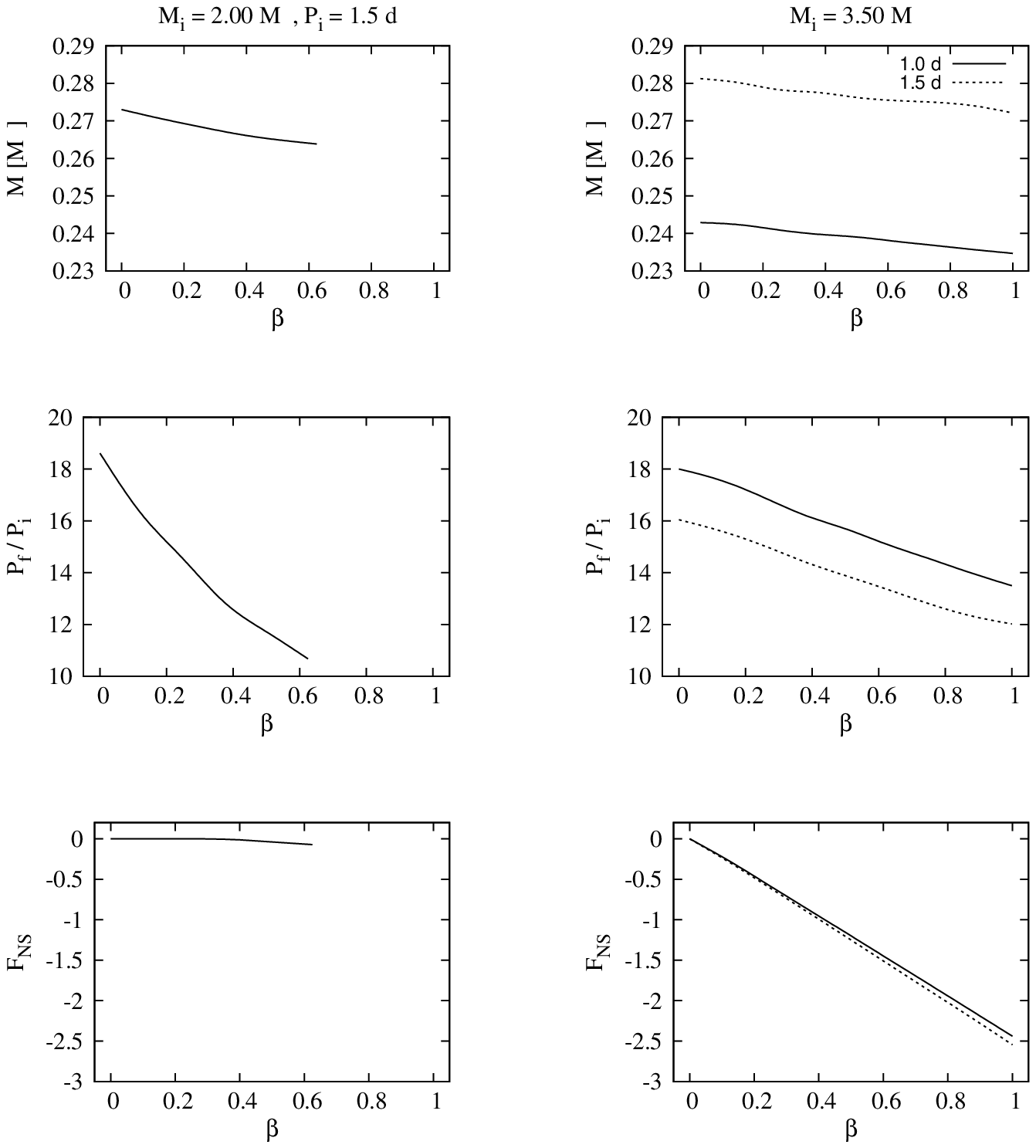}   
\caption{The final mass of the donor star (upper panels), the ratio of
  the final  to the  initial orbital period  (middle panels),  and the
  function $F_{\rm  NS}= M_{\rm NS}-(M_{\rm NS})_{i} +  \beta \big(M -
  M_{i}   \big)$  (defined   in   Eq.~\ref{eq:transf_mns},  see   also
  Fig.~\ref{fig:superf_mns})   (lower  panels)   as   a  function   of
  $\beta$.   Left   panels  correspond   to   systems  with   $(M_{\rm
    NS})_{i}=$~1.4~$M_{\odot}$      and     $M_{i}=$~2.00~$M_{\odot}$,
  $P_{i}=$~1.5~d,  while right  panels depict  the cases  of  the same
  initial  NS mass  and $M_{i}=$~3.50~$M_{\odot}$,  $P_{i}=$~1.0~d and
  1.5~d. For  these high mass values  of the donor star,  the range of
  initial periods  that produce a  helium WD is much  more restricted,
  which does not allow  for constructing surfaces like those presented
  in   Figs.~\ref{fig:superf_mdonor}-\ref{fig:superf_mns}.
\label{fig:graficos1}}
\end{figure*}

From the analysis  give above, we find that  changes in $\beta$ induce
smooth changes in the configuration  of the resulting CBSs. In view of
this fact we  have constructed surfaces to study  the behaviour of the
masses  of both stars  and the  final orbital  period as  functions of
$\beta$  and $\log_{10}{(P_{i}/days)}$.  As we  are interested  on the
formation of  helium WDs,  the surfaces cover  an ample region  of the
parameter space  only for the case  of donors with  low initial masses
(say  1.00  to 1.50~$M_{\odot}$);  for  higher  initial donor  masses,
surfaces are much  narrower.  These surfaces give a  direct insight on
the   dependence  of   evolutionary  sequences   with   the  parameter
$\beta$. In Fig.~\ref{fig:superf_mdonor} we show the mass of the donor
star remnant  as a  function of $\beta$  and $\log_{10}{(P_{i}/days)}$
for systems  with initial masses  $M_{i}/M_{\odot}$ of 1.00,  1.25 and
1.50 for  the donor star and  $(M_{\rm NS})_{i}=$~1.40~$M_{\odot}$ for
the NS component.  In Fig.~\ref{fig:superf_porbital} we show the ratio
of  the final  to  the initial  orbital  periods for  the same  models
included in Fig.~\ref{fig:superf_mdonor}.

In order to present our results  regarding the final NS mass, we found
it  useful to  make a  simple transformation.  In our  models  we have
assumed that

\begin{equation} 
\dot{M}_{\rm NS}= Min\big(\beta |\dot{M}|, \dot{M}_{Edd}\big).
\end{equation} 

If $\beta |\dot{M}| \le \dot{M}_{Edd}$ were fulfilled throughout the
entire evolution of the system, we may integrate it, finding that

\begin{equation} 
M_{\rm NS}-(M_{\rm NS})_{i}= -\beta \big( M - M_{i} \big),
\end{equation} 

where  $M$ and  $M_{\rm NS}$  stand for  the final  WD and  NS masses,
respectively.  So, we may define $F_{\rm NS}$ as

\begin{equation} \label{eq:transf_mns}
F_{\rm NS}= M_{\rm NS}-(M_{\rm NS})_{i} + \beta \big( M - M_{i} \big).
\end{equation} 

Clearly, if  $\dot{M}_{\rm NS} \le  \dot{M}_{Edd}$ is fulfilled
in   all   RLOFs,  then   $F_{\rm   NS}=   0$.   Thus,  $F_{\rm   NS}$
(Eq.~\ref{eq:transf_mns}) shows the effects due to the stages at which
$\dot{M}_{\rm NS} > \dot{M}_{Edd}$ forcing a supplementary mass
loss  rate  from  the  system  apart  from  the  $(1-\beta)  |\dot{M}|$
contribution. In Fig.~\ref{fig:superf_mns} we show the surface defined
by   $F_{\rm  NS}$   for  the   same   set  of   models  included   in
Figs.~\ref{fig:superf_mdonor}-\ref{fig:superf_porbital}.

In    view   of    the    fact   that    the    surfaces   shown    in
Figs.~\ref{fig:superf_mdonor}-\ref{fig:superf_mns} are very smooth, it
is useful to represent them  by a suitable function $F(x,y)$. We found
it adequate to consider the quotient of polynomials given by

\begin{equation}
F(x,y)= \frac{ C_1 + C_2 x + C_3 y + C_4 x^2 + C_5 y^2 + C_6 x y}
             { 1 + C_7 x + C_8 y }, \label{eq:fit}
\end{equation}

where  we assign  $x=  \beta$ and  $y= \log_{10}{(P_{i}/days)}$.   The
coefficients  $C_{i}, i=  1, \cdots,  8$ are  found by  standard least
squares     method     and     the     results    are     given     in
Table~\ref{table:fits}. These fits  allow for an immediate calculation
of the  final orbital  period and donor  mass values.  To  compute the
final NS  mass we  have firstly  to compute donor  star mass  and then
apply Eq.~(\ref{eq:transf_mns}).

These  fits provide  a useful  description  of the  dependence of  the
characteristics  of  these  systems  as  a  function  of  $\beta$  and
$\log_{10}{(P_{i}/days)}$.  While  these  surfaces correspond  to  the
cases of initial masses of $M_{i}/M_{\odot}=$ 1.00, 1.25, and 1.50 and
$(M_{\rm  NS})_{i}=$~1.40~$M_{\odot}$, it should  be stressed  that in
defining the surfaces we do not have to compute a large number of time
consuming binary evolutionary  sequences. If necessary, this technique
may be extended  to other values of donor and  NS masses employing the
results  given  in  Tables~\ref{table:mns0.80}-\ref{table:mns1.40}  of
Appendix~\ref{app_grid}.

As stated above,  for the case of more massive  donor stars, the range
of initial periods for which CBSs evolve to produce helium WDs is much
narrower, making it impossible  to construct surfaces similar to those
already presented.   In Fig.~\ref{fig:graficos1} we show  2D plots for
selected  systems, showing  the  dependence of  the  final masses  and
period of  these systems  as functions of  $\beta$.  The  behaviour of
these quantities is similar to that found for the case of less massive
donor stars.

\section{Discussion and Conclusions} \label{sec:discu}

In this paper  we have computed the evolution  of close binary systems
(CBSs)  composed of  a normal,  solar  composition, donor  star and  a
neutron star (NS) companion. The  range of masses and periods has been
chosen in order to study CBSs  that evolve to open, helium white dwarf
(WD)-millisecond  pulsar (MSP)  pairs or  to ultracompact  systems. In
order  to  compute the  orbital  evolution  of  these system  we  have
considered that  the NS is able  to accrete a $\beta$  fraction of the
mass coming  from the donor  star. We assumed  an upper limit  for the
accretion rate of the NS, imposed by the Eddington accretion rate (see
\S~\ref{sec:intro}).  In previous works  we have studied the evolution
of  CBSs varying  the initial  configuration, defined  by  the orbital
period, and the  masses of the donor and NS. In  this work we explored
the evolution of CBSs for a variety of values of $\beta$.

While       our      model       results       are      given       in
Tables~\ref{table:mns0.80}-\ref{table:mns1.40},   in  some  favourable
cases we  have been able to  construct surfaces for the  final mass of
the donor star, the orbital period and  the mass of the NS by means of
the function $F_{\rm NS}$  given in Eq.~(\ref{eq:transf_mns}). Also we
presented fits (Eq.~\ref{eq:fit}  and Table~\ref{table:fits}) to these
surfaces  that  allow   for  a  fast  evaluation  of   the  final  CBS
configuration as a function of  $P_{i}$ and $\beta$. It is clear that,
in the case  of the systems that evolve to  an open configuration, the
mass  of  the  resulting  WD  is  barely sensitive  to  the  value  of
$\beta$. The final orbital period,  in most cases (for a given initial
configuration)        exhibits        moderate       changes        of
approximately~$25\%$. However,  in some particular  cases, changes are
even larger  than 100\%  (see, e.g., in  Table~\ref{table:mns1.00} the
case  of  $M_i=2.50~M_{\odot}$,  $(M_{NS})_i=  1.00~M_{\odot}$,  $P_i=
1.50$~d).  As  expected, the most  sensitive quantity is the  final NS
mass.

\begin{centering} \begin{table*}
\caption{\label{table:shapiro-delay-data}  The  close  binary  systems
  composed of a millisecond pulsar and  a low mass WD for which it has
  been possible  to detect  the Shapiro delay  effect and  measure the
  masses of both components.  All these systems belong to the Galactic
  plane population.  From  left to right, the Table  presents the name
  of  the pulsar,  its  spin period,  the  WD and  pulsar masses,  the
  orbital period and the relevant reference respectively.}
\begin{tabular}{lccccl}
\hline \hline
Name & $P_p$ & $M_{WD}$ & $M_{\rm NS}$ & $P$ & Reference \\
     & [ms]  & [\msol]  & [\msol]  & [d] &          \\
\hline 
 PSR~J0437-4715 & 5.757  & $0.236 \pm 0.017$       & $1.58 \pm 0.18$        & 5.741  & van Straten et al. (2001)   \\
 PSR~J1614+2230 & 3.15   & $0.500\pm 0.006$	   & $1.97 \pm 0.04$	    & 8.687 & Demorest et al. (2010)	   \\
 PSR~J1713+0747 & 4.57   & $0.28 \pm 0.03$	   & $1.3 \pm 0.2$	    & 67.825 & Splaver et al. (2005)	   \\
 PSR~B1855+09	& 5.362  & $0.258^{+0.028}_{-0.016}$ & $1.50^{+0.26}_{-0.14}$   & 12.327 & Kaspi, Taylor \& Ryba (1994) \\
 PSR~J1909-3744 & 2.947  & $0.2038 \pm 0.0022$     & $1.438 \pm 0.024$      & 1.533  & Jacoby et al. (2005)	   \\
\hline \hline \end{tabular} \end{table*} \end{centering}

The  grid  presented in  this  paper should  be  useful  to study  the
characteristics of CBSs  composed of low mass WDs  and MSPs. There are
five systems  of this kind  (see Table~\ref{table:shapiro-delay-data})
in which  the masses  of the components  have been measured  with high
precision  thanks  to  the  detection of  the  relativistic Shapiro
delay in pulsar  timing (see Taylor \& Weisberg  1989 and references
therein).  Let us  now discuss the possibility of  inferring the value
of $\beta$ from the data available of these systems.

Looking for a  progenitor configuration of a CBS is a  way to test the
theory  of  CBS evolution.  If  correct, we  should  be  able to  find
plausible  initial  configurations.  The  most  sensitive quantity  to
changes of $\beta$ is the final NS  mass.  In order to recycle a NS we
do need a minimum amount of  mass to be accreted (see Cook, Shapiro \&
Teukolsky  1994). The  exact  value  of such  minimum  amount of  mass
depends  on the  initial rotation  rate  of the  NS and  on the  still
uncertain  cold nuclear  matter equation  of state.  This uncertainty,
together with the  unknown initial mass and spin of  the NS prevent us
from employing the  observational data available on the  NS to further
constrain the     parameter     space     for     the     initial
configuration.  Remarkably,  WD  properties  are barely  dependent  on
$\beta$  while in  most  cases  the orbital  period  shows a  moderate
dependence  with  this  parameter.  Thus,  it  is  very  difficult  to
determine $\beta$ by means of  evolutionary studies. This would be the
case, even if  the masses of the components of  the systems were known
far more accurately.

In  this  work,  as usual,  we  considered  the  value of  $\beta$  as
constant.  Of  course, this  may not  be the actual  case. If  so, our
$\beta$ could be considered as a kind of effective mean value.  In
any  case, our  results  indicate that  if  we are  interested in  the
evolution of the donor star  moving on a circular orbit, considering a
fixed $\beta$ is justified.   Including modeling of physical processes
that may modulate the accretion rate onto the NS (e.g.  magnetic field
evolution) has a minor effect  on the results presented in this paper.
Thus, these improvements are not warranted in this context.

Another  interesting issue  is the  fact that  existing models  do not
reproduce the  orbital period distribution  of binary MSP  well. Also,
the proposition  that low mass X-ray binaries  (LMXBs) are progenitors
of binary MSPs has the difficulty that the birthrates of both types of
systems do not match each  other.  Very recently, Hurley et al. (2010)
performed an  exhaustive study  of the birthrate  of LMXBs  and binary
MSPs and their orbital  period distributions. These authors considered
the  formation  of  NSs  that  become  MSPs  as  due  to  core-colapse
supernovae    and    accretion-induced    collapse   of    oxygen/neon
WDs.  Figs.~4-6 of  Hurley  et  al. (2010)  show  the theoretical  and
observed  orbital  period distribution  of  binary  MSPs. Despite  the
amount  of  details  considered  in  that study,  they  found  a  poor
agreement. In  this paper we have  found some dependence  of the final
orbital period with the value  of $\beta$.  Considering this effect in
the context of the orbital period distribution of binary MSPs may help
to  bring   theoretical  predictions  closer   to  observations.  This
possibility certaily deserves a detailed study.

The authors  want to thank to  our anonymous referee  for his/her very
useful and constructive reports that  guided us to largely improve the
original version of this paper.


\appendix \section{The grid of models}\label{app_grid}

In Tables~\ref{table:mns0.80}-\ref{table:mns1.40}  we present the main
results of our  calculations. Each Table corresponds to  a fixed value of
the initial  mass of the NS $(M_{\rm NS})_{i}$  and gives the final
masses of  the donor and NS  ($M$ and $M_{\rm  NS}$ respectively), and
the final orbital period $P_f$ for each initial configuration (defined by
the  initial mass  of the donor  star $M_{i}$, the  initial orbital
period $P_{i}$, and $\beta$).  We do not present results corresponding
to  initial donor  masses of  0.50, 0.65,  and 0.80~\msun  because the
corresponding systems  produce donor star with final masses of 
0.05~\msun or eventually the donor star overfill its Roche lobe on the
ZAMS.

\onecolumn
\begin{landscape}
\tiny

\begin{center}
\begin{longtable}{|c|ccc|ccc|ccc|ccc|ccc|ccc|ccc|}
\caption{Results corresponding to the case of an initial neutron star mass of $(M_{NS})_i =
0.80~M_{\odot}$. Each  horizontal block corresponds to a  value of the
initial mass  of the donor star, $M_i$.  In each block it  can be seen
the dependence  of our results  with the initial orbital  period $P_i$
(horizontally) and the value  of $\beta$ (vertically). For each $P_i$,
we give  the final  orbital period, $P_{f}$  (in days), and  the final
masses of the  donor and accreting neutron star,  $M$ and $M_{NS}$ (in
solar units),  respectively.  "$\dot M$ divergent"  corresponds to the
case in  which $|\dot{M}|  > 10^{-5}~M_{\odot}/year$; "$\dot M$  in ZAMS"
corresponds to  objects that  overfill their Roche  lobe on  the ZAMS;
"$log T_c \ge  8$" corresponds to objects that  ignite helium in their
cores; "$M_{NS}\ge 2.5$" indicates that the mass of the NS exceeds the
limit value of 2.5~\msol.  Numbers in italics corresponds to the value
of  the  orbital  period  at   13~Gyr,  for  systems  that  evolve  to
ultracompact configurations  in which the  final mass of  the donor
star is larger than 0.15~\msol. \label{table:mns0.80} }\\
\hline \hline \\[-2ex]   
   \multicolumn{1}{|c|}{} &   
   \multicolumn{3}{c|}{$P_i =$ 0.50~d} &  
   \multicolumn{3}{c|}{$P_i =$ 0.75~d} & 
   \multicolumn{3}{c|}{$P_i =$ 1.00~d} &  
   \multicolumn{3}{c|}{$P_i =$ 1.50~d} &
   \multicolumn{3}{c|}{$P_i =$ 3.00~d} &       
   \multicolumn{3}{c|}{$P_i =$ 6.00~d} &   
   \multicolumn{3}{c|}{$P_i =$12.00~d} \\ 
\hline \hline
$\beta$ &  $P_f$ & $M$ & $M_{NS}$ & 
	   $P_f$ & $M$ & $M_{NS}$ & 
           $P_f$ & $M$ & $M_{NS}$ &
	   $P_f$ & $M$ & $M_{NS}$ &
	   $P_f$ & $M$ & $M_{NS}$ & 
	   $P_f$ & $M$ & $M_{NS}$ &
           $P_f$ & $M$ & $M_{NS}$ \\[0.5ex]
\multicolumn{1}{c|}{}  & $[d]$ & $[M_{\odot}]$ & $[M_{\odot}]$ & 
   $[d]$ & $[M_{\odot}]$ & $[M_{\odot}]$ & 
   $[d]$ & $[M_{\odot}]$ & $[M_{\odot}]$ & 
   $[d]$ & $[M_{\odot}]$ & $[M_{\odot}]$ & 
   $[d]$ & $[M_{\odot}]$ & $[M_{\odot}]$ & 
   $[d]$ & $[M_{\odot}]$ & $[M_{\odot}]$ & 
   $[d]$ & $[M_{\odot}]$ & $[M_{\odot}]$ \\[0.5ex] 
 	   
\hline \hline \\[-1.8ex]
\endfirsthead

\multicolumn{22}{c}{-- Continued} \\[0.5ex]

\hline \hline \\[-2ex]   
   \multicolumn{1}{|c|}{} &   
   \multicolumn{3}{c|}{$P_i =$ 0.50~d} &  
   \multicolumn{3}{c|}{$P_i =$ 0.75~d} & 
   \multicolumn{3}{c|}{$P_i =$ 1.00~d} &  
   \multicolumn{3}{c|}{$P_i =$ 1.50~d} &
   \multicolumn{3}{c|}{$P_i =$ 3.00~d} &       
   \multicolumn{3}{c|}{$P_i =$ 6.00~d} &   
   \multicolumn{3}{c|}{$P_i =$12.00~d} \\ 
\hline \hline
$\beta$ & $P_f$ & $M$ & $M_{NS}$ & 
	   $P_f$ & $M$ & $M_{NS}$ & 
           $P_f$ & $M$ & $M_{NS}$ &
	   $P_f$ & $M$ & $M_{NS}$ &
	   $P_f$ & $M$ & $M_{NS}$ & 
	   $P_f$ & $M$ & $M_{NS}$ &
           $P_f$ & $M$ & $M_{NS}$ \\[0.5ex] 
\multicolumn{1}{c|}{}  & $[d]$ & $[M_{\odot}]$ & $[M_{\odot}]$ & 
   $[d]$ & $[M_{\odot}]$ & $[M_{\odot}]$ & 
   $[d]$ & $[M_{\odot}]$ & $[M_{\odot}]$ & 
   $[d]$ & $[M_{\odot}]$ & $[M_{\odot}]$ & 
   $[d]$ & $[M_{\odot}]$ & $[M_{\odot}]$ & 
   $[d]$ & $[M_{\odot}]$ & $[M_{\odot}]$ & 
   $[d]$ & $[M_{\odot}]$ & $[M_{\odot}]$ \\[0.5ex] 
 	   
\hline \hline \\[-1.8ex]

\endhead

\multicolumn{22}{l}{{Continued on next page \ldots}} \\
\endfoot

\endlastfoot
\multicolumn{22}{|c|}{$M_i =$ 1.00~$M_{\odot}$} \\
\hline
0.00 & 0.027 & 0.049 & 0.800 & 1.139 & 0.178 & 0.800 & 3.946 & 0.212 & 0.800 & 9.193 & 0.234 & 0.800 & 23.560 & 0.261 & 0.800 & 45.773 & 0.283 & 0.800 & 82.802 & 0.302 & 0.800 \\  
0.25 & 0.027 & 0.050 & 1.038 & 1.375 & 0.182 & 1.004 & 4.488 & 0.215 & 0.995 & 9.787 & 0.236 & 0.965 & 22.602 & 0.261 & 0.937 & 45.222 & 0.280 & 0.859 & 77.446 & 0.300 & 0.900 \\
0.50 & 0.028 & 0.050 & 1.275 & 1.516 & 0.184 & 1.207 & 4.492 & 0.218 & 1.174 & 9.408 & 0.235 & 1.132 & 21.524 & 0.260 & 1.084 & 40.365 & 0.279 & 1.023 & 72.826 & 0.298 & 0.999 \\
0.75 & 0.026 & 0.050 & 1.513 & 1.722 & 0.186 & 1.408 & 4.610 & 0.216 & 1.327 & 9.447 & 0.235 & 1.245 & 20.473 & 0.258 & 1.200 & 38.099 & 0.277 & 1.135 & 68.628 & 0.296 & 1.052 \\
1.00 & 0.026 & 0.050 & 1.740 & 1.790 & 0.186 & 1.581 & 4.400 & 0.218 & 1.461 & 8.963 & 0.233 & 1.372 & 19.306 & 0.257 & 1.306 & 36.058 & 0.275 & 1.208 & 65.365 & 0.294 & 1.128 \\
\hline \hline
\multicolumn{22}{|c|}{$M_i =$ 1.25~$M_{\odot}$} \\
\hline
0.00 & 0.040 & 0.050 & 0.800 & 0.012 & 0.050 & 0.800 & 3.381 & 0.209 & 0.800 &\multicolumn{3}{c|}{$\dot M$ divergent}& & & & & & & & & \\
\hline \hline
\end{longtable}
\end{center}
\newpage
\begin{center}
\begin{longtable}{|c|ccc|ccc|ccc|ccc|ccc|ccc|ccc|}
\caption{Results corresponding to the case of an initial neutron star mass of $(M_{NS})_i=1.00~M_{\odot}$. 
Its structure is the same as that of Table~\ref{table:mns0.80}. \label{table:mns1.00} }\\
\hline \hline \\[-2ex]   
   \multicolumn{1}{|c|}{} &   
   \multicolumn{3}{c|}{$P_i =$ 0.50~d} &  
   \multicolumn{3}{c|}{$P_i =$ 0.75~d} & 
   \multicolumn{3}{c|}{$P_i =$ 1.00~d} &  
   \multicolumn{3}{c|}{$P_i =$ 1.50~d} &
   \multicolumn{3}{c|}{$P_i =$ 3.00~d} &       
   \multicolumn{3}{c|}{$P_i =$ 6.00~d} &   
   \multicolumn{3}{c|}{$P_i =$12.00~d} \\ 
\hline \hline
$\beta$ & $P_f$ & $M$ & $M_{NS}$ & 
	   $P_f$ & $M$ & $M_{NS}$ & 
           $P_f$ & $M$ & $M_{NS}$ &
	   $P_f$ & $M$ & $M_{NS}$ &
	   $P_f$ & $M$ & $M_{NS}$ & 
	   $P_f$ & $M$ & $M_{NS}$ &
           $P_f$ & $M$ & $M_{NS}$ \\[0.5ex]
\multicolumn{1}{c|}{}  & $[d]$ & $[M_{\odot}]$ & $[M_{\odot}]$ & 
   $[d]$ & $[M_{\odot}]$ & $[M_{\odot}]$ & 
   $[d]$ & $[M_{\odot}]$ & $[M_{\odot}]$ & 
   $[d]$ & $[M_{\odot}]$ & $[M_{\odot}]$ & 
   $[d]$ & $[M_{\odot}]$ & $[M_{\odot}]$ & 
   $[d]$ & $[M_{\odot}]$ & $[M_{\odot}]$ & 
   $[d]$ & $[M_{\odot}]$ & $[M_{\odot}]$ \\[0.5ex]  	   
 	   
\hline \hline \\[-1.8ex]
\endfirsthead

\multicolumn{22}{c}{-- Continued} \\[0.5ex]

\hline \hline \\[-2ex]   
   \multicolumn{1}{|c|}{} &   
   \multicolumn{3}{c|}{$P_i =$ 0.50~d} &  
   \multicolumn{3}{c|}{$P_i =$ 0.75~d} & 
   \multicolumn{3}{c|}{$P_i =$ 1.00~d} &  
   \multicolumn{3}{c|}{$P_i =$ 1.50~d} &
   \multicolumn{3}{c|}{$P_i =$ 3.00~d} &       
   \multicolumn{3}{c|}{$P_i =$ 6.00~d} &   
   \multicolumn{3}{c|}{$P_i =$12.00~d} \\ 
\hline \hline
$\beta$ & $P_f$ & $M$ & $M_{NS}$ & 
	   $P_f$ & $M$ & $M_{NS}$ & 
           $P_f$ & $M$ & $M_{NS}$ &
	   $P_f$ & $M$ & $M_{NS}$ &
	   $P_f$ & $M$ & $M_{NS}$ & 
	   $P_f$ & $M$ & $M_{NS}$ &
           $P_f$ & $M$ & $M_{NS}$ \\[0.5ex] 
\multicolumn{1}{c|}{}  & $[d]$ & $[M_{\odot}]$ & $[M_{\odot}]$ & 
   $[d]$ & $[M_{\odot}]$ & $[M_{\odot}]$ & 
   $[d]$ & $[M_{\odot}]$ & $[M_{\odot}]$ & 
   $[d]$ & $[M_{\odot}]$ & $[M_{\odot}]$ & 
   $[d]$ & $[M_{\odot}]$ & $[M_{\odot}]$ & 
   $[d]$ & $[M_{\odot}]$ & $[M_{\odot}]$ & 
   $[d]$ & $[M_{\odot}]$ & $[M_{\odot}]$ \\[0.5ex] 
 	   
\hline \hline \\[-1.8ex]

\endhead

\multicolumn{22}{l}{{Continued on next page \ldots}} \\
\endfoot

\endlastfoot
\multicolumn{22}{|c|}{$M_i =$ 1.00~$M_{\odot}$} \\
\hline
0.00 & 0.030 & 0.050 & 1.000 & 1.488 & 0.183 & 1.000 & 5.328 & 0.221 & 1.000 & 12.716 & 0.240 & 1.000 & 32.000 & 0.270 & 1.000 & 62.460 & 0.292 & 1.000 & 111.815 & 0.314 & 1.000 \\
0.25 & 0.027 & 0.050 & 1.238 & 1.991 & 0.191 & 1.202 & 5.496 & 0.223 & 1.194 & 12.484 & 0.242 & 1.189 & 30.707 & 0.270 & 1.182 & 59.116 & 0.290 & 1.175 & 103.935 & 0.312 & 1.152 \\
0.50 & 0.030 & 0.050 & 1.475 & 1.794 & 0.188 & 1.405 & 5.606 & 0.223 & 1.388 & 12.327 & 0.241 & 1.376 & 29.055 & 0.267 & 1.355 & 54.776 & 0.288 & 1.321 &  97.655 & 0.308 & 1.281 \\
0.75 & 0.027 & 0.050 & 1.713 & 2.209 & 0.196 & 1.602 & 5.598 & 0.224 & 1.581 & 11.942 & 0.241 & 1.550 & 27.499 & 0.266 & 1.504 & 51.372 & 0.286 & 1.450 &  91.058 & 0.307 & 1.378 \\ 
1.00 & 0.027 & 0.050 & 1.950 & 2.292 & 0.197 & 1.802 & 5.649 & 0.222 & 1.754 & 11.650 & 0.238 & 1.701 & 25.985 & 0.263 & 1.641 & 48.590 & 0.284 & 1.565 &  87.662 & 0.304 & 1.465 \\
\hline \hline
\multicolumn{22}{|c|}{$M_i =$ 1.25~$M_{\odot}$} \\
\hline
0.00 & 0.040 & 0.050 & 1.000 & 0.011 & 0.050 & 1.000 & 4.993 & 0.223 & 1.000 & 14.705 & 0.246 & 1.000 &\multicolumn{3}{c|}{$\dot M$ divergent}& 71.873 & 0.296 & 1.000 & 126.245& 0.319 & 1.000 \\
\hline \hline
\end{longtable}
\end{center}

\newpage
\begin{center}
\begin{longtable}{|c|ccc|ccc|ccc|ccc|ccc|ccc|ccc|}
\caption{Results corresponding to the case of an initial neutron star mass of $(M_{NS})_i =
1.20~M_{\odot}$. Its structure is the same as that of Table~\ref{table:mns0.80}. \label{table:mns1.20} }\\
\hline \hline \\[-2ex]   
   \multicolumn{1}{|c|}{} &   
   \multicolumn{3}{c|}{$P_i =$ 0.50~d} &  
   \multicolumn{3}{c|}{$P_i =$ 0.75~d} & 
   \multicolumn{3}{c|}{$P_i =$ 1.00~d} &  
   \multicolumn{3}{c|}{$P_i =$ 1.50~d} &
   \multicolumn{3}{c|}{$P_i =$ 3.00~d} &       
   \multicolumn{3}{c|}{$P_i =$ 6.00~d} &   
   \multicolumn{3}{c|}{$P_i =$12.00~d} \\ 
\hline \hline
$\beta$ & $P_f$ & $M$ & $M_{NS}$ & 
	   $P_f$ & $M$ & $M_{NS}$ & 
           $P_f$ & $M$ & $M_{NS}$ &
	   $P_f$ & $M$ & $M_{NS}$ &
	   $P_f$ & $M$ & $M_{NS}$ & 
	   $P_f$ & $M$ & $M_{NS}$ &
           $P_f$ & $M$ & $M_{NS}$ \\[0.5ex]
\multicolumn{1}{c|}{}  & $[d]$ & $[M_{\odot}]$ & $[M_{\odot}]$ & 
   $[d]$ & $[M_{\odot}]$ & $[M_{\odot}]$ & 
   $[d]$ & $[M_{\odot}]$ & $[M_{\odot}]$ & 
   $[d]$ & $[M_{\odot}]$ & $[M_{\odot}]$ & 
   $[d]$ & $[M_{\odot}]$ & $[M_{\odot}]$ & 
   $[d]$ & $[M_{\odot}]$ & $[M_{\odot}]$ & 
   $[d]$ & $[M_{\odot}]$ & $[M_{\odot}]$ \\[0.5ex] 
   	   
\hline \hline \\[-1.8ex]
\endfirsthead

\multicolumn{22}{c}{-- Continued} \\[0.5ex]

\hline \hline \\[-2ex]   
   \multicolumn{1}{|c|}{} &   
   \multicolumn{3}{c|}{$P_i =$ 0.50~d} &  
   \multicolumn{3}{c|}{$P_i =$ 0.75~d} & 
   \multicolumn{3}{c|}{$P_i =$ 1.00~d} &  
   \multicolumn{3}{c|}{$P_i =$ 1.50~d} &
   \multicolumn{3}{c|}{$P_i =$ 3.00~d} &       
   \multicolumn{3}{c|}{$P_i =$ 6.00~d} &   
   \multicolumn{3}{c|}{$P_i =$12.00~d} \\ 
\hline \hline
$\beta$ & $P_f$ & $M$ & $M_{NS}$ & 
	   $P_f$ & $M$ & $M_{NS}$ & 
           $P_f$ & $M$ & $M_{NS}$ &
	   $P_f$ & $M$ & $M_{NS}$ &
	   $P_f$ & $M$ & $M_{NS}$ & 
	   $P_f$ & $M$ & $M_{NS}$ &
           $P_f$ & $M$ & $M_{NS}$ \\[0.5ex] 
\multicolumn{1}{c|}{}  & $[d]$ & $[M_{\odot}]$ & $[M_{\odot}]$ & 
   $[d]$ & $[M_{\odot}]$ & $[M_{\odot}]$ & 
   $[d]$ & $[M_{\odot}]$ & $[M_{\odot}]$ & 
   $[d]$ & $[M_{\odot}]$ & $[M_{\odot}]$ & 
   $[d]$ & $[M_{\odot}]$ & $[M_{\odot}]$ & 
   $[d]$ & $[M_{\odot}]$ & $[M_{\odot}]$ & 
   $[d]$ & $[M_{\odot}]$ & $[M_{\odot}]$ \\[0.5ex] 
 	   
\hline \hline \\[-1.8ex]

\endhead

\multicolumn{22}{l}{{Continued on next page \ldots}} \\
\endfoot

\endlastfoot
\multicolumn{22}{|c|}{$M_i =$ 1.00~$M_{\odot}$} \\
\hline
0.00 & 0.032 & 0.050 & 1.200 & 1.868 & 0.190 & 1.200 & 6.800 & 0.227 & 1.200 & 15.801 & 0.246 & 1.200 & 39.491 & 0.276 & 1.200 & 76.043 & 0.299 & 1.200 & 135.766& 0.320 & 1.200 \\
0.25 & 0.029 & 0.050 & 1.438 & 2.437 & 0.196 & 1.400 & 6.943 & 0.230 & 1.392 & 15.248 & 0.248 & 1.387 & 36.988 & 0.274 & 1.381 &\multicolumn{3}{c|}{$\dot M$ divergent}&125.088& 0.318 & 1.368 \\ 
0.50 & 0.032 & 0.050 & 1.675 & 2.056 & 0.191 & 1.603 & 6.648 & 0.228 & 1.585 & 14.538 & 0.246 & 1.575 & 34.881 & 0.272 & 1.563 & 74.913 & 0.298 & 1.750 & 117.062& 0.315 & 1.522 \\
0.75 & 0.030 & 0.050 & 2.113 & 2.550 & 0.200 & 1.799 & 6.580 & 0.228 & 1.778 & 14.095 & 0.245 & 1.764 & 32.584 & 0.272 & 1.736 & 61.401 & 0.292 & 1.701 & 109.045& 0.313 & 1.642 \\
1.00 & 0.030 & 0.050 & 2.350 & 2.199 & 0.194 & 2.004 & 6.589 & 0.226 & 1.972 & 13.729 & 0.243 & 1.943 & 30.827 & 0.269 & 1.901 & 57.699 & 0.290 & 1.831 & 103.518& 0.311 & 1.737 \\
\hline \hline
\multicolumn{22}{|c|}{$M_i =$ 1.25~$M_{\odot}$} \\
\hline
0.00 & 0.040 & 0.050 & 1.200 & {\it 0.218} & 0.157 & 1.200 & 6.437 & 0.228 & 1.200 & 18.630 & 0.255 & 1.200 & 48.508 & 0.283 & 1.200 & 93.566 & 0.306 & 1.200 & 165.170& 0.329 & 1.200 \\
\hline \hline
\end{longtable}
\end{center}

\newpage
\begin{center}
\begin{longtable}{|c|ccc|ccc|ccc|ccc|ccc|ccc|ccc|}
\caption{Results corresponding to the case of an initial neutron star mass of $(M_{NS})_i =
1.40~M_{\odot}$, but for the case of $\Delta \beta = 0.125$. 
Its structure is the same as that of Table~\ref{table:mns0.80}.
\label{table:mns1.40} }\\
\hline \hline \\[-2ex]   
   \multicolumn{1}{|c|}{} &   
   \multicolumn{3}{c|}{$P_i =$ 0.50~d} &  
   \multicolumn{3}{c|}{$P_i =$ 0.75~d} & 
   \multicolumn{3}{c|}{$P_i =$ 1.00~d} &  
   \multicolumn{3}{c|}{$P_i =$ 1.50~d} &
   \multicolumn{3}{c|}{$P_i =$ 3.00~d} &       
   \multicolumn{3}{c|}{$P_i =$ 6.00~d} &   
   \multicolumn{3}{c|}{$P_i =$12.00~d} \\ 
\hline \hline
$\beta$ &  $P_f$ & $M$ & $M_{NS}$ & 
	   $P_f$ & $M$ & $M_{NS}$ & 
           $P_f$ & $M$ & $M_{NS}$ &
	   $P_f$ & $M$ & $M_{NS}$ &
	   $P_f$ & $M$ & $M_{NS}$ &
	   $P_f$ & $M$ & $M_{NS}$ & 
	   $P_f$ & $M$ & $M_{NS}$ \\[0.5ex]
\multicolumn{1}{c|}{}  & $[d]$ & $[M_{\odot}]$ & $[M_{\odot}]$ & 
   $[d]$ & $[M_{\odot}]$ & $[M_{\odot}]$ & 
   $[d]$ & $[M_{\odot}]$ & $[M_{\odot}]$ & 
   $[d]$ & $[M_{\odot}]$ & $[M_{\odot}]$ & 
   $[d]$ & $[M_{\odot}]$ & $[M_{\odot}]$ & 
   $[d]$ & $[M_{\odot}]$ & $[M_{\odot}]$ & 
   $[d]$ & $[M_{\odot}]$ & $[M_{\odot}]$ \\[0.5ex]

\hline \hline \\[-1.8ex]
\endfirsthead

\multicolumn{22}{c}{-- Continued} \\[0.5ex]

\hline \hline \\[-2ex]   
   \multicolumn{1}{|c|}{} &   
   \multicolumn{3}{c|}{$P_i =$ 0.50~d} &  
   \multicolumn{3}{c|}{$P_i =$ 0.75~d} & 
   \multicolumn{3}{c|}{$P_i =$ 1.00~d} &  
   \multicolumn{3}{c|}{$P_i =$ 1.50~d} &
   \multicolumn{3}{c|}{$P_i =$ 3.00~d} &       
   \multicolumn{3}{c|}{$P_i =$ 6.00~d} &   
   \multicolumn{3}{c|}{$P_i =$12.00~d} \\ 
\hline \hline
$\beta$ &  $P_f$ & $M$ & $M_{NS}$ & 
	   $P_f$ & $M$ & $M_{NS}$ & 
           $P_f$ & $M$ & $M_{NS}$ &
	   $P_f$ & $M$ & $M_{NS}$ &
	   $P_f$ & $M$ & $M_{NS}$ & 
	   $P_f$ & $M$ & $M_{NS}$ &
           $P_f$ & $M$ & $M_{NS}$ \\[0.5ex] 
\multicolumn{1}{c|}{}  & $[d]$ & $[M_{\odot}]$ & $[M_{\odot}]$ & 
   $[d]$ & $[M_{\odot}]$ & $[M_{\odot}]$ & 
   $[d]$ & $[M_{\odot}]$ & $[M_{\odot}]$ & 
   $[d]$ & $[M_{\odot}]$ & $[M_{\odot}]$ & 
   $[d]$ & $[M_{\odot}]$ & $[M_{\odot}]$ & 
   $[d]$ & $[M_{\odot}]$ & $[M_{\odot}]$ & 
   $[d]$ & $[M_{\odot}]$ & $[M_{\odot}]$ \\[0.5ex] 
 	   
\hline \hline \\[-1.8ex]

\endhead

\multicolumn{22}{l}{{Continued on next page \ldots}} \\
\endfoot

\endlastfoot

\multicolumn{22}{|c|}{$M_i =$ 1.00~$M_{\odot}$} \\
\hline
0.000 & 0.030 & 0.049 & 1.400 & 2.676 & 0.202 & 1.400 & 8.428 & 0.234 & 1.400 & 18.154 & 0.256 & 1.400 & 45.317 & 0.281 & 1.400 & 85.966 & 0.305 & 1.400 & 153.780 & 0.326 & 1.400\\
0.125 & 0.030 & 0.049 & 1.519 & 2.733 & 0.203 & 1.500 & 8.360 & 0.234 & 1.496 & 17.747 & 0.255 & 1.493 & 43.820 & 0.280 & 1.490 & 83.531 & 0.303 & 1.487 & 147.790 & 0.325 & 1.484\\
0.250 & 0.030 & 0.049 & 1.638 & 2.763 & 0.201 & 1.599 & 8.265 & 0.234 & 1.591 & 17.320 & 0.254 & 1.586 & 42.377 & 0.279 & 1.580 & 79.671 & 0.302 & 1.574 & 142.231 & 0.323 & 1.569\\
0.375 & 0.031 & 0.050 & 1.756 & 2.790 & 0.202 & 1.699 & 8.142 & 0.234 & 1.687 & 16.887 & 0.254 & 1.680 & 40.906 & 0.278 & 1.670 & 77.468 & 0.300 & 1.662 & 136.886 & 0.322 & 1.653\\
0.500 & 0.031 & 0.050 & 1.876 & 2.807 & 0.202 & 1.798 & 8.022 & 0.233 & 1.782 & 16.519 & 0.253 & 1.773 & 39.656 & 0.277 & 1.760 & 74.703 & 0.299 & 1.749 & 131.840 & 0.320 & 1.734\\
0.625 & 0.031 & 0.049 & 1.994 & 2.816 & 0.202 & 1.897 & 7.913 & 0.233 & 1.878 & 16.152 & 0.252 & 1.867 & 38.265 & 0.277 & 1.851 & 71.349 & 0.299 & 1.834 & 127.280 & 0.319 & 1.809\\
0.750 & 0.031 & 0.049 & 2.113 & 2.895 & 0.201 & 1.996 & 7.804 & 0.233 & 1.974 & 15.791 & 0.252 & 1.960 & 37.205 & 0.275 & 1.942 & 69.687 & 0.296 & 1.917 & 123.357 & 0.318 & 1.873\\
\hline
\hline
\end{longtable}

\end{center}


\normalsize
\end{landscape}

\twocolumn
\section{The relation between the mass of the white dwarf and the final orbital period}\label{app_mwdp}

\begin{figure} 
\epsfysize=300pt  
\epsfbox{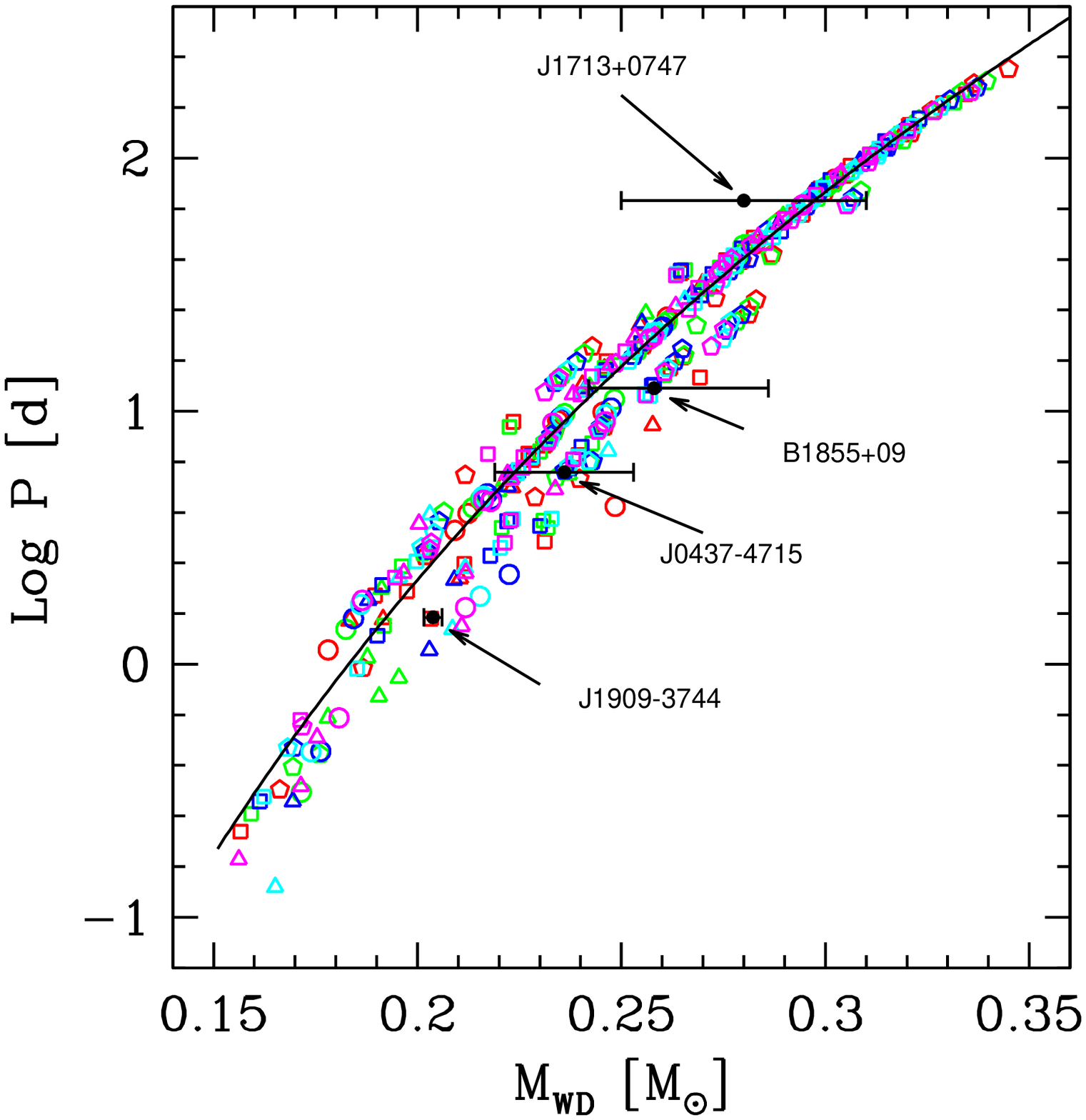} 
\caption{\label{fig:mwd_logp} The relation between  the mass of the WD
  and the  final orbital period.  We present with  circles, triangles,
  squares  and pentagons  the cases  of accreting  neutron  stars with
  initial   mass   values  of   0.80,   1.00,   1.20  and   1.40~\msun
  respectively.  The colors  red, green,  blue, sky  blue  and magenta
  correspond to the cases of  $\beta=$ 0.00, 0.25, 0.50, 0.75 and 1.00
  respectively.   Solid  line represents  the  fit  to the  $M_{WD}-P$
  relation $P=  2.6303 \times 10^{6}  \; (M_{WD}/M_{\odot})^{8.7078}\;
  d$  given in  De  Vito \&  Benvenuto  (2010).  We  have plotted,  in
  addition, data corresponding to four helium white dwarfs belonging 
  to close binary systems, companions of  millisecond pulsars, whose 
  masses are known.}
\end{figure}

The  mass of  WDs resulting  from the  evolution of  CBSs  satisfies a
relation with the  final orbital period of the  system.  This relation
has been studied previously by,  e.g., Rappaport et al. (1995), Tauris
\& Savonije (1999), Nelson,  Dubeau \& MacCannell (2004). Recently, we
have  examined  quantitatively the  predictions  made  by the  authors
previously  mentioned,  and  proposed  a  new relation  from  our  own
calculations (De Vito  \& Benvenuto 2010). In that  work we considered
the  evolution of donor  stars of  CBSs with  different masses  of the
accreting  NSs but a  fixed value  of $\beta$~(=0.5).  Now, we  add to
these calculations  the results  corresponding to different  values of
the parameter $\beta$. This is shown in Fig.~\ref{fig:mwd_logp}, where
we show  the results corresponding  to open systems and  to converging
systems in which the mass of  the donor star is larger than 0.15~\msun
(in these cases we  plot the value of the orbital period  at an age of
13~Gyr). We  observe from Fig.~\ref{fig:mwd_logp}  that the dispersion
in the  relation we have plotted  decreases as we  move towards larger
masses of the WD and orbital  periods. This is because in these cases,
the donor  star is  on the red  giant branch  at the beginning  of the
first mass  transfer episode, and then, the  core mass-radius relation
is well satisfied (Joss, Rappaport \& Lewin 1987).

The     WD    mass-orbital     period     relation    presented     in
Fig.~\ref{fig:mwd_logp} is  based on five  times more models  than that
shown in De  Vito \& Benvenuto (2010), which  makes the relation shown
here as  a more solid  result. In any  case we should remark  that the
analytical fitting  found in our  previous paper is in  nice agreement
with these new results.  A  variation of the value of $\beta$ produces
a motion of the  point at most in the size of  the symbols employed in
the figure.

\bsp \label{lastpage} 

\end{document}